\documentclass[%
 aip,
 jmp,%
 amsmath,amssymb,
 reprint,%
]{revtex4-2}

\usepackage{graphicx}% Include figure files
\usepackage{dcolumn}% Align table columns on decimal point
\usepackage{bm}% bold math
\usepackage{algorithm2e}
\usepackage{xcolor}
\usepackage{hyperref}
\usepackage{amssymb}
\usepackage{amsmath}

\newif\iffinal

\finalfalse

\iffinal
    \newcommand\kirill[1]{}
    \newcommand\nate[1]{}
    \newcommand\mike[1]{}
    \newcommand\marcus[1]{}
    \newcommand\ben[1]{}
    \newcommand\ian[1]{}
\else
    \newcommand\kirill[1]{{\color{red}[Kirill: #1]}}
    \newcommand\nate[1]{{\color{orange}[Nate: #1]}}
    \newcommand\mike[1]{{\color{blue}[Mike: #1]}}
    \newcommand\marcus[1]{{\color{cyan}[Marcus: #1]}}
    \newcommand\ben[1]{{\color{purple}[Ben: #1]}}
    \newcommand\ian[1]{{\color{teal}[Ian: #1}}
\fi

\newcommand{\mycomment}[1]{}

\begin{document}

\preprint{AIP/123-QED}

%\title[Sample title]{Sample Title:\\with Forced Linebreak\footnote{Error!}}% Force line breaks with \\
%\thanks{Footnote to title of article.}
%\title{Can One Hear the Shape of a Molecular Rotor?\\The Rotational Spectroscopy Inverse Problem
%}
\title{
%Isospectrality in rotational spectroscopy: does a rotational spectrum uniquely identify a molecule?
Twins in rotational spectroscopy: Does a rotational spectrum uniquely identify a molecule?
}

\author{Marcus Schwarting}
\affiliation{Department of Computer Science, University of Chicago, Chicago, IL 60637, USA}%Lines break automatically or can be forced with \\

\author{Nathan~A.~Seifert}%
\affiliation{ 
Department of Chemistry and Chemical \& Biomedical Engineering, University of New Haven, West Haven, CT 06516, USA
}%

\author{Michael~J.~Davis}
\affiliation{%
Chemical Sciences and Engineering Division, Argonne National Laboratory, Lemont, IL 60439, USA}%

\author{Ben Blaiszik}
\affiliation{Data Science and Learning Division, Argonne National Laboratory, Lemont, IL 60439, USA}

\author{Ian Foster}
\affiliation{Department of Computer Science, University of Chicago, Chicago, IL 60637, USA}
\affiliation{Data Science and Learning Division, Argonne National Laboratory, Lemont, IL 60439, USA}

\author{Kirill Prozument}
\affiliation{Chemical Sciences and Engineering Division, Argonne National Laboratory, Lemont, IL 60439, USA}

\date{\today}% It is always \today, today,

\begin{abstract}
%Up-to-date version of abstract, based on multiple iterations with Kirill and Mike.
Rotational spectroscopy is the most accurate method for determining structures of molecules in the gas phase.
It is often assumed that a rotational spectrum is a unique “fingerprint” of a molecule.
The availability of large molecular databases and the development of artificial intelligence methods for spectroscopy makes the testing of this assumption timely.
In this paper, we pose the determination of molecular structures from rotational spectra as an inverse problem.
Within this framework, we adopt a funnel-based approach to search for molecular twins, which are two or more molecules, which have similar rotational spectra but distinctly different molecular structures.
We demonstrate that there are twins within standard levels of computational accuracy by generating rotational constants for many molecules from several large molecular databases, indicating the inverse problem is ill-posed.
However, some twins can be distinguished by increasing the accuracy of the theoretical methods or by performing additional experiments.

\end{abstract}

\keywords{Inverse problems, rotational spectroscopy, isospectral geometry}%Use showkeys class option if keyword
\maketitle

%\begin{quotation}
%The ``lead paragraph'' is encapsulated with the \LaTeX\ 
%\verb+quotation+ environment and is formatted as a single paragraph before the first section heading. 
%(The \verb+quotation+ environment reverts to its usual meaning after the first sectioning command.) 
%Note that numbered references are allowed in the lead paragraph.
%
%The lead paragraph will only be found in an article being prepared for the journal \textit{Chaos}.
%\end{quotation}

\section{\label{sec:introduction}Introduction}

Pure rotational spectroscopy is a powerful spectroscopic technique in the microwave and millimeter-wave frequency ranges that can reveal detailed structural and dynamical information about a molecule in the gas phase that is not obtainable with other spectroscopic techniques \cite{Wilson1968}.
The invention of broadband chirped-pulse Fourier transform microwave (CP-FTMW) spectroscopy (also called molecular rotational resonance spectroscopy) enabled fast acquisition of data over many GHz of spectral bandwidth with sub-MHz resolution and meaningful relative intensities of spectral lines \cite{Pate2008}.
Because CP-FTMW offers simultaneous quantitative detection of multiple species in the gas phase with isomer, conformer, and quantum state specificity, it has replaced or complemented the previous generations of microwave spectrometers in physical chemistry laboratories \cite{Pate2012, Dan2021, Ian2022, BB2023, Melanie2023}.
However, its potential remains largely untapped in analytical chemistry or industrial settings in part because assignment of unknown spectra and identifying the molecules that give rise to those spectra requires a trained spectroscopist \cite{Justin2024}.

Spectral assignment entails attributing experimentally observed spectral lines to transitions between quantum levels with known quantum numbers.
That assignment in rotational spectroscopy is based on a quantum mechanical model that adequately describes molecular rotation and intramolecular interactions \cite{Pate2023}.
Identifying the correct set of parameters in that model, such as the rotational constants, distortion constants, and electric quadrupole interaction constants, is a non-trivial task. Efforts to automate this task are underway \cite{seifert2015autofit, Dan2018, mccarthy2020molecule}, but
even when spectral assignment is complete and an experimental spectrum can be simulated by solving the forward problem, the chemical identity of a molecule often remains unknown or ambiguous.
Currently, the chemical identity is guessed and verified by calculating the molecular geometry by using \textit{ab initio} methods, solving the forward problem, and comparing the simulated and measured spectra.
The inverse problem in rotational spectroscopy is to identify a molecular geometry either from the set of rotational constants or from the spectrum itself \cite{Nate2020, mccarthy2020molecule, cheng2023reflection}.
A solution necessarily exists, but is it unique?
In this work we study the latter inverse problem, namely: can a rotational spectrum uniquely define a molecule?

%This work is split into sections providing background on inverse problems and rotational spectroscopy, our methods for assessing isospectrality by constructive and exhaustive methods, the results of our isospectral assessments, a discussion of the implications for rotational spectroscopy as an alternative identification technique, our conclusions, and future work. \nate{This last paragraph should be a bit like the abstract -- more concretely introduce our approach and our conclusions.}
The rest of this paper is as follows.
In Section~\ref{sec:background},
we provide background on inverse problems and isospectrality, %in a spectroscopic context%
and on how rotational spectra are analyzed in both the forward and inverse contexts.
Next in Section~\ref{sec:methods}, we introduce our constructive and exhaustive methods to assess the isospectral nature of rotational spectra.
We then present our results in Section~\ref{sec:results}, first for our constructed environments and then for the datasets we analyzed.
Finally, we discuss the implications of these findings when using rotational spectroscopy for sample identification, and consider future directions.
%Finally, we conclude that the simulated mapping from molecules to rotational spectra appears to be well-posed, while the experimental mapping from molecules to rotational spectra can remain ambiguous in some instances.

\section{\label{sec:background}Background}

We first provide a brief introduction to inverse problems and their relevance to spectroscopic analysis, and then review the current state-of-the-art in forward and inverse mapping approaches within rotational spectroscopy.

\subsection{\label{sec:bg_inv_probs}Inverse Problems and Isospectrality in Spectroscopy}
Inverse problems can be broadly defined as follows: For some deterministic forward process $f$ (e.g., a dynamical system, machine learning model inference, simulation, or experimental procedure), can one predict $f^{-1}$, i.e., the input associated with a specific output\cite{tarantola2005inverse}?
A natural extension to this question is whether such an inverse mapping from an output to an input is unique.
That property of $f$ is known as well-posedness (also called injectivity); a pair of inputs that result in the same output, and thus demonstrate that $f$ is not well posed, is known as an isospectral collision.
Many inverse problems are ill-posed; that is, solutions may be non-unique.
In 1966, Mark Kac famously described and explored the inverse problem ``can one hear the shape of a drum?'', which poses the question of whether an individual with perfect pitch (capable of accurately describing the entire set of frequencies associated with a sound) can uniquely identify the shape of a drum (defined as a membrane uniformly stretched across a topologically compact region $\Omega \subset \mathbb{R}^2$) by the sounds it produces \cite{kac1966can}.
For Kac's query, the forward mapping consisted of applying the Laplacian wave equation across an input surface $\Omega$ and identifying nontrivial normal modes through a  deterministic process, leading to a discrete series of ordered ``tones.''
The isospectrality problem of determining whether a given set of tones is unique to an input surface $\Omega$ is known to hold for convex surfaces, but counterexamples for concave surfaces have since been identified \cite{gordon1992one}.

Outside of acoustics, variations of Kac's original question have been explored in a variety of domains, including imaging \cite{bertero2021introduction}, signal processing \cite{dokmanic2011can}, photonics \cite{park2021hearing}, quantum mechanics \cite{pursey1986isometric}, and spectroscopy \cite{heilbronner1978spectral}.
Many such isospectrality problems in this area of research, commonly known as spectral geometry, remain unsolved or have been solved only under a set of strictly limiting constraints.
Furthermore, while Kac and others assume that $f$ may be perfectly observed, in practice whether or not $f$ is well-posed also relies on measurement precision.
Two distinct inputs may be distinguishable when measured at higher resolution, but not when measured at lower resolution.

Following the publication of H{\"u}ckel's molecular orbital (HMO) theory \cite{coulson1978huckel}, spectral geometry was first employed for chemical systems.
H{\"u}ckel presented a method to compute the molecular orbital $\vert \psi \rangle$ of $\pi$-conjugated systems from a simple linear combination of $2p_z$ atomic orbitals $\vert \phi_i \rangle$ with corresponding coefficients $\{ c_i \}$, written as $\vert \psi_i \rangle = c_1 \vert \phi_1 \rangle + c_2 \vert \phi_2 \rangle$.
Substituting the above form into the Schr{\"o}dinger equation, we may write the secular equation $(\boldsymbol{H}-E\boldsymbol{S})\overrightarrow{c}=0$ where $\boldsymbol{S}_{i,j}=\langle \phi_i \vert \phi_j \rangle$ is the overlap matrix and $H_{i,j}=\langle \phi_i \vert \boldsymbol{\hat{H}} \vert \phi_j \rangle$ is the Hamiltonian matrix.
Nontrivial eigenvalues from this secular formulation correspond to the respective atomic orbital energies of the system.
G{\"u}nthard and Primas showed how coordinated $\pi$-bonds in HMO theory could be represented concisely with a graph adjacency matrix \cite{gunthard1956zusammenhang} and considered whether distinct molecular graphs representing $\pi$-coordinated HMO systems would always have distinct sets of eigenvalues.
Collatz and Singowitz first identified isospectral collisions among simple graphs \cite{von1957spektren} and many chemically relevant isospectral collisions (or near collisions \cite{dias1996almost}) have since been pinpointed \cite{herndon1975isospectral,van2003graphs,hosoya2016chemistry}.
Other molecular representations have also been considered, with Schrier \cite{schrier2020can} using a supervised machine learning approach to demonstrate that a set of constitutional isomers of acyclic alkanes cannot be perfectly distinguished by using the Coulomb matrix eigenvalues \cite{rupp2012fast} as a descriptor.

Inverse and isospectrality problems are increasingly relevant for spectroscopy and analytical chemistry.
The analytical power of a spectroscopic technique lies primarily in the degree to which molecules can be uniquely distinguished from one another.
When a spectroscopic technique yields results that lead to structural ambiguity, it is common for experimenters to use additional spectroscopic techniques to resolve remaining ambiguity.
Even after collecting multiple measurements, some structural ambiguity may remain (such as distinguishing between enantiomers).
Nuclear magnetic resonance (NMR) and infrared (IR) spectroscopy are popular analytical techniques because they can rapidly resolve most structural ambiguities to identify a sample.
These techniques come with the added bonus that measured spectra can be immediately interpreted to yield structural insights.
A significant portion of the inverse problem for small molecules can be performed by human experts or heuristic-based scripts, and interpreting such spectra is a topic covered in most undergraduate chemistry curricula.
While rotational spectroscopy greatly surpasses IR spectroscopy in its precision for determining molecular structure in the gas phase, IR spectroscopy is far easier to interpret.
%Structural insights are difficult to obtain from a rotational spectrum alone.
Obtaining structural insights from a rotational spectrum alone is not straightforward.

\subsection{\label{sec:bg_spec_analysis}
Forward and Inverse Mapping in Rotational Spectroscopy}
\begin{figure}[b]
\includegraphics[width=0.95\textwidth]{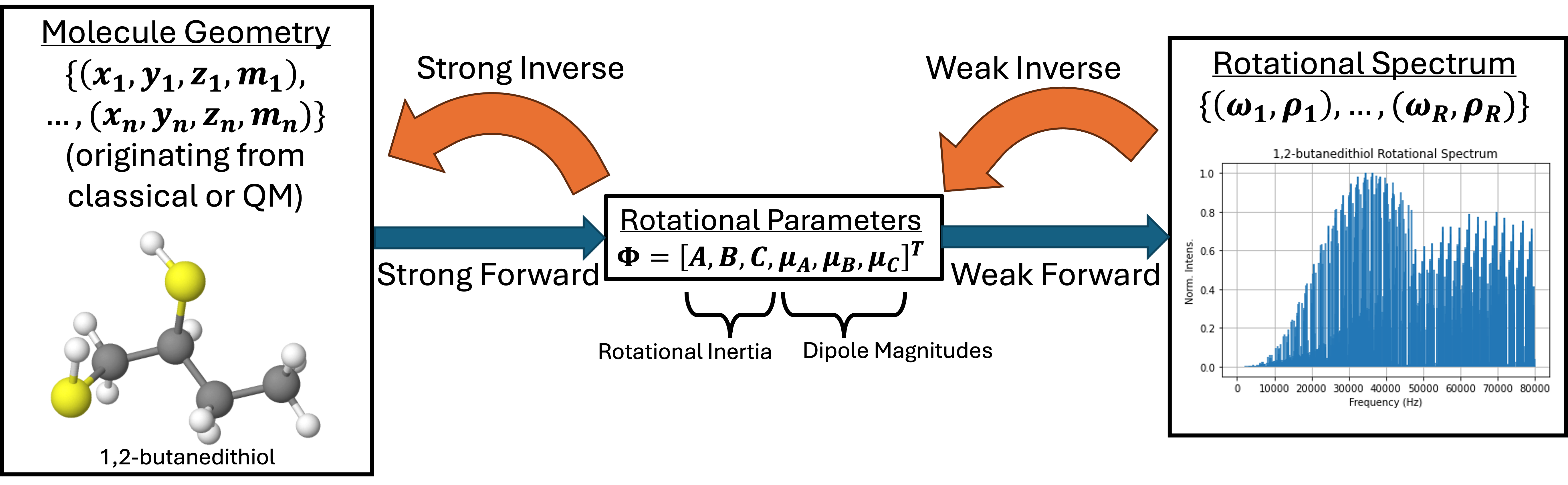}
\caption{\label{fig:inverse_diagram} Diagram showing forward and inverse mapping in rotational spectroscopy.}
\end{figure}

%Rotational spectra can be analyzed %
Molecules can be related to their rotational spectra via either a forward mapping (from molecular geometry to spectrum) or an inverse mapping (from spectrum to molecular descriptors).
The forward mapping occurs in two steps: the molecule geometry to a set of rotational and dipole constants, and this set of constants to the rotational spectrum \cite{kroto1975molecular}.
We term the former the \textit{strong} forward problem and the latter the \textit{weak} forward problem.
Likewise, the mapping from rotational spectrum to the set of constants is termed the \textit{weak} inverse problem, and mapping from the set of constants to the molecule geometry is termed the \textit{strong} inverse problem.
\autoref{fig:inverse_diagram} illustrates the forward and strong/weak inverse problems. 

\subsubsection{Twins and Isospectral Collisions}
%\kirill{
Uniqueness of the solution for the weak and the strong inverse problems may be expressed with the help of some additional definitions.
We assert that two molecules constitute an isospectral collision if they are \textit{indistinguishable} by rotational spectroscopy; that is, if their rotational lines cannot be resolved by a rotational spectrometer.
For the microwave region, this normally means that their frequencies are within $\sim$10~kHz.
%However we can also define a looser constraint: we also molecules \textit{twins} if their best available simulated rotational constants are identical within the 1\% uncertainty.
We can also define a looser constraint: we call molecules \textit{twins} if their experimentally measured spectra are distinct from one another, but in the event they are both present in an experimental spectrum, it is unclear which spectrum can be attributed to which molecule. %CONTINUE WORKSHOPPING THIS
This distinction arises primarily from the aleatoric uncertainty inherent in aligning simulated molecule constants with experimentally identified molecule constants, a problem that has been studied by \citet{lee2020bayesian}
%This distinction arises primarily from the accuracy of nowadays structure calculations that can be performed in a reasonable time to assist molecular identification.
If a set of indistinguishable molecules can be identified, the inverse problem (weak and strong) is ill-posed.
For twin molecules, the weak inverse problem is well-posed (that is, no distinct sets of rotational and dipole constants map to the same spectrum), but the strong inverse problem is generally ill-posed.
However, the strong inverse problem can be made well-posed for twin molecules with additional information.
This additional information may be obtained by measuring the dipole moment directions derived from relative line intensities, performing higher fidelity simulations of the molecular structure, measuring isotopically-substituted species, collecting Stark or nutation measurements of the dipole moment, or observing intramolecular interactions identified from line splittings.
None of these approaches strictly require measurements from a separate spectroscopic technique.
%}

%Collisions: two molecules result in spectra that are completely indistinguishable (eg. enantiomers)
%Twins: two molecules result in spectra that are distinct, but one cannot tell which molecule is which merely by looking at the spectra.

Tackling an inverse problem starts with first defining the forward problem.
When considering the forward problem, we assume that an optimized molecule geometry is already available via some classical or quantum mechanical technique.
\citet{lee2020bayesian} present an expected margin of error in DFT-derived rotational and dipole constants compared to experiment, an important consideration when evaluating isospectral constraints.
Assuming a pre-computed geometry, we briefly review the strong forward mapping for deriving the rotational and dipole constants of a molecular geometry.
(For more information on the formulation of the weak forward mapping, see \citet{gordon1992one} and \citet{kroto1975molecular}.)
Finally, we describe current efforts towards efficiently solving both the weak and strong inverse mapping problems.

\subsubsection{Strong Forward Mapping}
A conformer can be defined by six variables: three rotational constants ($A,B,C$) and three corresponding dipole constants ($\mu_A,\mu_B,\mu_C$).
Taking the conformer geometry, derived via force field, ab-initio, or wave function approaches, as a starting point, suppose a conformer is defined as $\{(m_1,x_1,y_1,z_1), ...,  (m_n,x_n,y_n,z_n)\}$.
Then we may define an inertia matrix as
$$\boldsymbol{A} = \begin{bmatrix} 
  I_{x,x} & I_{x,y} & I_{x,z}\\
  I_{y,x} & I_{y,y} & I_{y,z}\\
  I_{z,x} & I_{z,y} & I_{z,z}
\end{bmatrix}
$$
where each element represents a moment of inertia along a pair of Cartesian axes.
On-diagonal elements are calculated as
$$I_{x,x} = \sum_{i=0}^n m_i((y_i-\bar{y})^2 + (z_i - \bar{z})^2)$$
and off-diagonal elements are calculated as
$$I_{x,y} = -\sum_{i=0}^n m_i (x_i - \bar{x})(y_i - \bar{y})$$
which are normalized to the center of mass $(\bar{x},\bar{y},\bar{z})$, calculated as
$$\bar{x} = \frac{\sum_{i=0}^n m_i x_i}{\bar{m}};\; \bar{m}=\sum_{i=0}^n m_i.$$
Rotational constants $(A,B,C)$ are calculated as 
$B_K=h/(8\pi^2 I_{B_K})$ 
%$B_K=\frac{h}{8\pi^2 I_{B_K}}$ 
(where $h$ is Planck's constant) by using eigenvalues $I_{B_K}$ of the matrix $\boldsymbol{A}$, ordered as $A\geq B \geq C$.
From a dipole vector oriented in Cartesian space $(\mu_x,\mu_y,\mu_z)$, the rotational dipoles $(\mu_A,\mu_B,\mu_C)$ can be calculated by using a change-of-basis with the rotational eigenmatrix.
An important quantity for measuring the degree of asymmetry of a molecule is Ray's asymmetry parameter 
$\kappa = (2B-A-C)/(A-C)$,
%$\kappa = \frac{2B-A-C}{A-C}$, 
where $\kappa=-1$ implies a perfectly prolate rotor and $\kappa=1$ implies a perfectly oblate rotor.
Based on a collection of rotational constants for a diverse set of roughly 400 molecules aggregated by \citet{hellwege1967landolt}, \citet{silbey1988preponderance} first observed that most measured molecules were highly prolate.
Silbey et al. then derived an equation for the probability distribution of $\kappa$ based on the construction of a random collection of point masses.
%$$P(\kappa)=\frac{\sqrt{2}}{(1-\kappa)^{3/2}} \biggr[ \biggr( \frac{3-\kappa}{8} \biggr) \biggr[ \ln\biggr( 1+\sqrt{\frac{1-\kappa}{2}}\biggr) -\ln\biggr( 1 - \sqrt{\frac{1-\kappa}{2}} \biggr) - \frac{1}{2}\sqrt{\frac{1-\kappa}{2}} \biggr] \biggr].$$
As far as we know, this equation is the only attempt to describe the shape distribution of all possible molecular rotors.

\subsubsection{Inverse Mapping}
The weak inverse problem of mapping rotational spectra to a set of rotational and dipole constants remains challenging.
However, several semi-automated packages are available to aid researchers.
In instances where a subset of rotational transition peaks can be labelled reliably, SPFIT and PGOPHER use a linear least squares procedure to  determine the rotational constants accurately \cite{pickett1991fitting,western2017pgopher}.
When a smaller set of transition peaks is available and clear bounds on rotational constants are known, AUTOFIT can  determine accurately a set of rotational constants for multiple conformers through a brute-force grid approach \cite{seifert2015autofit}, which has since been scaled to high-performance computing systems \cite{di2021testing}.
When no transition peaks can be  assigned manually, the RAARR package \cite{yeh2019automated} can mark certain trends in peaks by type (scaffolds) by using heuristics pointed out by \citet{cooke2013decoding}.
However, RAARR requires that strong a-type and b-type peaks be present in order to construct such trends, and many molecules do not exhibit these peaks.
%\kirill{
A spectrum of a single molecular carrier may be assigned a set of rotational constants and electric quadrupole constants by using the RAINet artificial neural network \cite{Dan2018}.
RAINet is trained on simulated spectra of several classes of molecular rotors (linear, symmetric top, asymmetric a-type, b-type, c-type, with different nuclear spins).
Classification and regression take about 200 $\mu$s regardless of the spectral complexity.
However, RAINet does not discern spectra from multiple carriers unless it is trained on such mixtures.
Other work considers how various distance metrics can be used to measure the space between experimentally observed and computationally proposed spectra \cite{seifert2021computational, seifert2022computational, seifert2023computational}.

The strong inverse problem of mapping from rotational and dipole constants to molecular structures is a more daunting challenge.
One intuitive approach uses a lookup table to map, for a large set of computed rotational spectra, directly to a molecule identity, thus obviating the need for rotational and dipole constants to be determined \cite{carroll2021high}.
This lookup approach has successfully been applied to a complex mixture of benzene gasses \cite{mccarthy2020exhaustive}.
\citet{mccarthy2020molecule} use a two-step probabilistic deep learning framework to reveal structural information from a set of rotational and dipole constants.
First a neural network is employed to predict the largest Coulomb matrix eigenvalues, then a second (probabilistic) neural network is applied to these eigenvalues to predict structural information, including the SMILES string and the presence of various functional groups.
However, as \citet{schrier2020can} points out for a set of acyclic alkanes, and as McCarthy and Lee also determine, the lossy Coulomb matrix eigenvalue representation cannot uniquely predict structural information for a molecule.
Finally, recent work from \citet{cheng2023reflection} shows how a diffusion-based model can derive structural insights from a set of labelled rotational constants for a parent species and corresponding isotopomeric species by using Kraitchman's equations \cite{kraitchman1953determination} and learning a positive or negative assignment for atomic coordinates.

\section{Methods}\label{sec:methods}

%\ian{Would be good to explain what these are methods FOR. Some insight is (in my view) missing.
%I realize that this is stated in the abstract (at least it will be if you add mention of the constructive approach), but I think that it should be spelt out here in more detail.
%As I understand things, the goal is to identify isospectral molecules. 
%You take two approaches to this problem,
%In the first constructive approach, you  start with one  molecule, and construct an isospectral partner by adding atoms. That is implicit in what follows, but the reader could benefit from stating this explicitly.
%I also suspect that defining term constrained and unconstrained at the start would be helpful.
%In the second search-based approach, you search large datasets ...
%}

%In this section we first introduce constructed environments at both a constrained and unconstrained extreme, and describe a procedure for identifying near-isospectral collisions.
In this section we first introduce two constructed environments which generate structures in a constrained and an unconstrained setting.
The constrained and unconstrained environments represent two extrema for structural enumeration, with the space of chemically feasible molecular structures situated somewhere between these two extremes.
We then describe a funnel process for identifying potential isospectral collisions within large datasets of molecules.

\subsection{Isospectrality by Construction}
Suppose we derive rotational constants $(A,B,C)$ for an initial molecule with a given relaxed geometry $\mathcal{M}$.
For this fixed structure $\mathcal{M}$, we wish to construct a distinct molecule $\mathcal{M}'$ that is isospectral to $\mathcal{M}$ through an iterative addition of atoms.
Note that $\mathcal{M}'$ must not be equivalent to $\mathcal{M}$ via translation, rotation, or reflection, but must possess rotational constants $(A',B',C')$ that are indistinguishable from $(A,B,C)$.
We define \textit{indistinguishable} here to mean that constants are similar to within measurable experimental error.
For small molecules that exhibit only a small number of measurable peaks between 2 and 18 GHz (a common frequency range for structure determination studies that employ microwave spectroscopy \cite{perez2013broadband}), the threshold for experimental error may be greater than for larger molecules that exhibit many measurable peaks at fixed intervals based on peak type.
We assume that, regardless of the molecules in question, \textit{indistinguishable} implies that frequencies of the pairs of observed spectral lines are within $\sim 10$ kHz. 

When one adds an atom and corresponding bonds to create a new molecule, the geometry must be re-optimized, which affects the Cartesian coordinates of all atoms in the system.
%In other words, adding a point mass to an existing structure would not guarantee that the re-relaxed structure has rotational constants $\leq10$ kHz of the unrelaxed structure.
In other words, adding a point mass to an existing structure would not guarantee that the re-relaxed structure maintains pairs of observed spectral lines within $\sim 10$ kHz of the unrelaxed structure.
Therefore, such a constructive process could be computationally expensive and may be unlikely to yield the precise collision of rotational constants desired.
In contrast to real chemical space, we consider two examples at opposite extremes which do not require relaxation: a constrained environment and an unconstrained environment.
The general process of adding atoms to form new valid molecular geometries falls somewhere between these two extremes.

%\subsubsection{Construction within a Constrained Environment}
%TODO: DEFINE CONSTRAINED AND UNCONSTRAINED

In a constrained environment, structures are defined by a single bond length, a single bond angle, and a single unitary atomic mass.
Suppose we begin with a unitary point mass at the origin of $\mathbb{R}^3$ and are permitted iteratively to add unitary point masses only at points that are of unit length away from the existing structure along the x-, y-, or z-axis, with no point masses being placed on top of one another.
Such a structure will eventually resemble a square lattice.
%Appendix \ref{append:constrained_env} elaborates on the construction of such lattices.
Supplementary Information Section I elaborates on the construction of such lattices.
%%%%%%%%%%%%%%%%%%%%%%%%%%%%%%%%%%%%%%%%%%%%%%

%\subsubsection{Construction within an Unconstrained Environment}
Now we consider the opposite extreme: in an unconstrained environment, structures are defined by any arbitrary bond length, any arbitrary bond angle, and any possible atomic mass.
Suppose that we begin with a structure $\mathcal{S}$ of $n$ unitary point masses, $n>2$, with masses $m_1,...,m_n \in \mathbb{R}^+$ and each with a different position in Cartesian space:
$$\mathcal{S}=\{(x_1,y_1,z_1),...,(x_n,y_n,z_n) \vert (x_i,y_i,z_i) \in \mathbb{R}^3 \}$$.
Suppose further that we can continuously alter the Cartesian coordinates of $\mathcal{S}$ to optimize towards a target $\boldsymbol{I_C}=(I_{x,x},I_{x,y},I_{x,z},I_{y,y},I_{y,z},I_{z,z})$.
It is straightforward to identify isospectral collisions in this unconstrained environment.
For example, the structure $\mathcal{S}_1 = \{ (0,1,0),(1,0,0),(0,-1,0) \}$ with corresponding masses $(1,2,1)$ is an isospectral collision with $\mathcal{S}_2 = \{ (0,1,0),(\sqrt{3},0,0),(0,-1,0) \}$ with corresponding masses $(1,1,1)$.
We can further justify that, for an arbitrary structure, there are infinitely many distinct structures that are isospectral to the initial spectrum.
Finally, we can show that the process of optimizing towards an isospectral collision is nonconvex, with many degrees of freedom allowing for many possible collisions.
%Appendix \ref{append:unconstrained_env} elaborates on the construction and optimization of these unconstrained structures.
Supplementary Information Section II elaborates on the construction and optimization of these unconstrained structures.

\subsection{Isospectrality by Molecule Assessment}
Rather than attempting to construct geometries from scratch, especially when these geometries may not resemble valid molecules, we also employ large datasets of relaxed molecular geometries to evaluate the potential collisions.
Considering relaxed geometries also allows us to evaluate other physical constrains beyond rotational constants that affect the resulting rotational spectrum.
%Considering relaxed geometries also
%allows us to evaluate beyond rotational constants to other factors that affect the resulting
%rotational spectrum.

\subsubsection{The Isospectral Funnel}
\begin{figure}[b]
\includegraphics[width=0.65\textwidth]{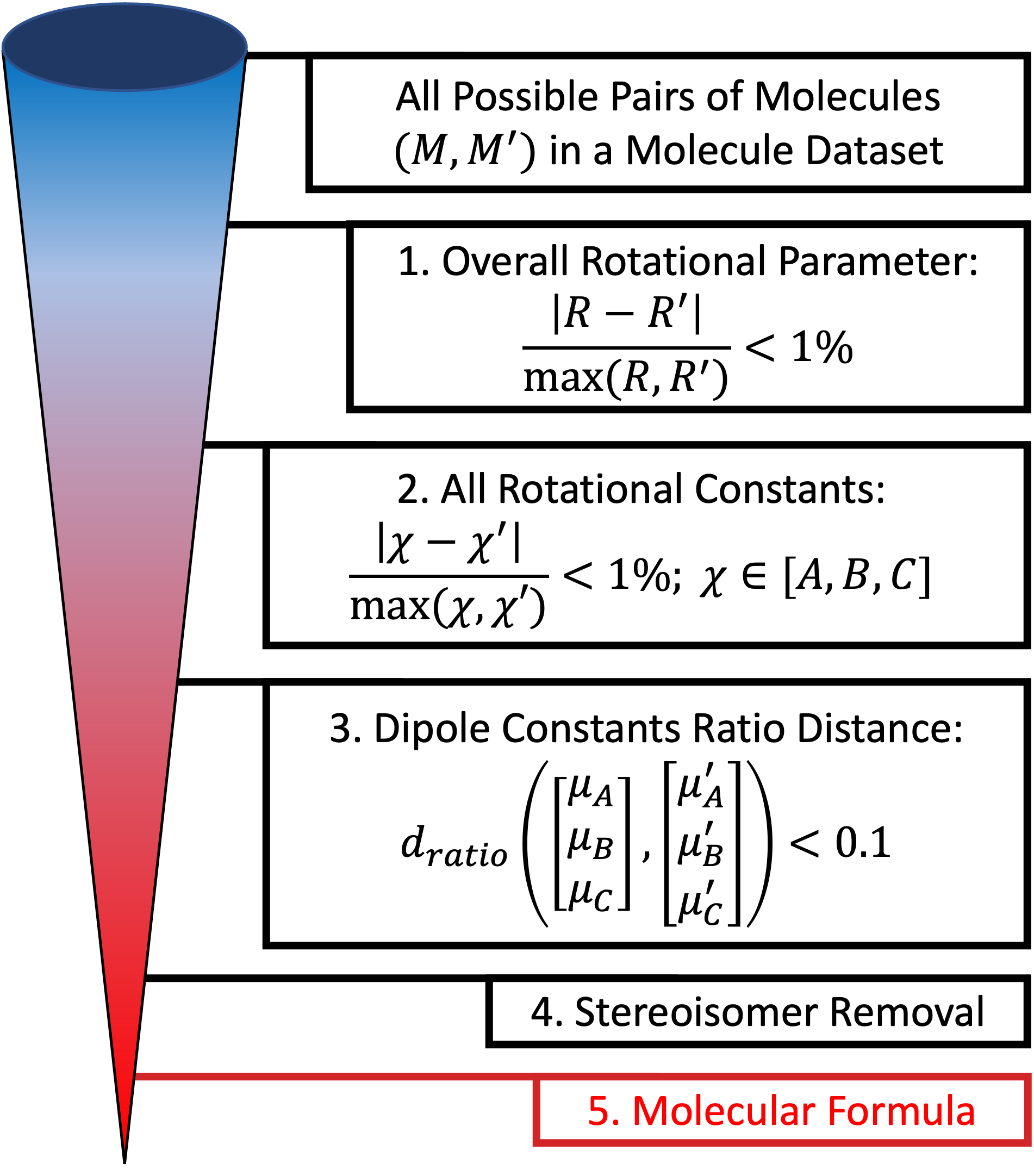}
\caption{\label{fig:funnel_diagram} Funnel diagram for evaluating possible spectral twins in a dataset of molecule geometries.}
\end{figure}

When evaluating possible collisions, we use a funnel-based approach as shown in \autoref{fig:funnel_diagram}.
Each successive step applies a more rigorous but also more expensive test to remove possible molecule pairs from consideration.
With the initial comparisons of rotational constants, we find that $>$90\% of possible molecule pairs may be removed from consideration.
While comparing rotational and dipole constants between a single molecule pair is computationally inexpensive, the number of considered molecule pairs is large enough to motivate our funnel-based approach to reduce computational overhead.

Starting from a large set of molecule geometries, we begin by enumerating all possible pairs of molecules $(M,M')$.
We then reduce the set of possible pairs based on an overall rotational parameter $R=\sqrt{A^2 + B^2 + C^2}$.
We perform a percentage comparison by dividing $\vert R-R'\vert$ by $\max(R,R')$, and retain all pairs where the percent difference is $<$1\%.
When molecules are placed in order by an ascending value of $R$, a percentage comparison is a computationally efficient method to eliminate a significant fraction of possible pairs.
Next we compare individual rotational constants, where pairs must satisfy a percent difference of $<$1\% for all of $(A,B,C)$.
Another significant fraction of possible pairs can be eliminated in this fashion.

We next perform a similar comparison of dipole ratios, but with several adjustments and edge cases.
%First, let $r_{A} = \frac{\mu_A^2}{(\mu_A')^2};\; r_{B} = \frac{\mu_B^2}{(\mu_B')^2};\; r_{C} = \frac{\mu_C^2}{(\mu_C')^2}$.
First, let $r_{A} = \left(\frac{\mu_A}{\mu_A'}\right)^2;\; r_{B} = \left(\frac{\mu_B}{\mu_B'}\right)^2;\; r_{C} = \left(\frac{\mu_C}{\mu_C'}\right)^2$.
Next, we normalize these squared dipole ratios according to $\bar{r_A}=\frac{r_A}{\max(r_A,r_B,r_C)};\;\bar{r_B}=\frac{r_B}{\max(r_A,r_B,r_C)};\;\bar{r_C}=\frac{r_C}{\max(r_A,r_B,r_C)}$.
We then compare pairs of ratios as $\rho_{A,B}=|\bar{r_A} - \bar{r_B}|;\;\rho_{B,C}=|\bar{r_B} - \bar{r_C}|;\;\rho_{C,A}=|\bar{r_C} - \bar{r_A}|$.
If $\max(\rho_{A,B},\rho_{B,C},\rho_{C,A})$ is less than the specified tolerance, then the pair satisfies our dipole constraint.
We specify an absolute tolerance of 0.1 to these dipole ratio comparisons.
This metric arises from the assumption that the abundances of the species, $n$, are not known.
Therefore, because the CP-FTMW signal is proportional to $n\mu^2$\cite{Pate2008}, molecules $\mathcal{M}$ and $\mathcal{M}'$ cannot be distinguished by the ratios $\left(\frac{\mu_\chi}{\mu_\chi'}\right)^2$.
However, the difference in relative intensities of the $a$-type and $b$-type transitions within each spectrum, for example, are observable and expressed here through $\rho_{A,B}$.

Since experimental measurements of rotational spectra are extremely accurate in the frequency domain (with peak frequencies arising solely from the contribution of rotational constants) and are less accurate in the intensity domain (with peak intensities arising from the contribution of both rotational and dipole constants, and further complicated by non-equilibrium experimental conditions \cite{puzzarini2013rotational}, and imperfections in the apparatus calibration), we assign the tolerances of rotational and dipole constants according to this difference in simulation and detection accuracy.
There are several edge cases with respect to this approach of comparing dipole magnitudes.
First, we presume that any dipole magnitude $<0.05 D$ cannot be readily measured experimentally.
We also presume that any dipole magnitude $>0.1 D$ can be measured experimentally.
For a pair of molecules, suppose $\mu_\chi<0.05$ and $\mu_\chi'>0.1$.
Then this pair cannot be a set of twins, because the first species exhibits no $\chi$-type peaks, while the second species exhibits $\chi$-type peaks.
If instead $\mu_\chi<0.05$ and $0.05<\mu_\chi'<0.1$,
then we employ the standard ratio approach to compare these dipoles.
If rather $\mu_\chi<0.05$ and $\mu_\chi'<0.05$ (which would occur for a planar or near-planar molecule),
then we omit the $r$ values that include this dipole component, leaving the remaining value of $\rho$ which does not include the $\chi$ dipole component as the only source of comparison for the tolerance.
Finally, suppose $\mu_{\chi_1}<0.05;\;\mu_{\chi_2}<0.05$ and $\mu_{\chi_1}'<0.05;\;\mu_{\chi_2}'<0.05$ (which would occur for a linear or near-linear molecule).
Then, so long as the remaining dipole component is measurable for both species, the pair of molecules are within tolerance by default.

%Next, we compare molecules according to the hyperfine splitting that a spectrum would exhibit.
%Since the degree of hyperfine splitting arising from the interaction of the electric quadrupole moments of $^{14}$N atoms with the gradient of the molecular electric field is associated primarily with the number of nitrogen atoms \cite{knauer1976nitrogen}, we simply require that pairs must both contain the same number of nitrogen molecules.
Next, we remove molecule pairs which are stereoisomers of one another.
We use the MolVS package \cite{bento2020open} to remove stereoisomer pairs, and further convert structures to canonical SMILES strings which are agnostic to chirality \cite{o2012towards}.
Finally, we provide an optional comparison based on molecular formula.
While a molecular formula cannot be immediately determined from a rotational spectrum alone (as this is part of the strong inverse problem), it can be easily assessed by using mass spectrometry,\cite{mass_spec2008} which can be run in conjunction with rotational spectroscopy.

The funnel we have so far defined considers only a single conformer of a molecule without isotopic substitution.
In the case where multiple relaxed geometry conformers (and corresponding relative energies) are available per molecule, the same funnel process described above can occur by using the lowest-energy conformer of each pair, but with a final step that requires a less rigorous alignment for pairs of available higher energy conformer geometries.
Similarly, %a final%
an additional step in the funnel might consider spectral differences caused by isotopic substitution.
Rejecting pairs based on isotopic substitution assumes that the experimental signal-to-noise ratio (SNR) exceeds 300:1 in the case of $^{13} C$ substitution, 1500:1 in the case of $^{18} O$ substitution, and roughly 18000:1 in the case of deuterium substitution (assuming natural abundance without enrichment).
While a majority of instruments can capture $^{13} C$ substitution without difficulty, most instruments cannot easily capture deuterium substitution at natural abundance.
Since experimental SNR can vary widely by instrument, pre-sets, and measurement time, we do not include differences arising from the presence of isotopomers in rotational spectra as a source of further disambiguation, although we acknowledge that it could be used as such.
The same may be said for other measurable effects such as hyperfine interaction and internal rotation.

\subsubsection{Molecule Datasets}
We consider several molecule datasets spanning a range of molecule size, diversity, and geometry fidelity: see \autoref{tab:molecule_datasets}.
\textbf{QM9} \cite{ramakrishnan2014quantum} comprises $\sim$1.33$\times 10^5$ chemically valid structures with up to nine (C,O,N,F) heavy atoms (and implicit H atoms), providing for each a DFT-optimized geometry (using B3LYP/6-31G(2df,p)) and corresponding Cartesian-oriented dipole vector.
\textbf{QM7x} \cite{hoja2021qm7} includes 6950 chemically valid structures containing up to seven (C,N,O,S,Cl) heavy atoms (with implicit H atoms).
Unlike QM9, QM7x provides multiple DFT-optimized geometries and a Cartesian-oriented dipole vector for each molecule (using PBE0+MBD), for a total of
$4.03 \times 10^6$ DFT-optimized conformers. Thus each unique molecule in QM7x has an average of $\sim$580 distinct conformers.

Enumerating the conformational diversity of a set of large molecules by using high-fidelity DFT is computationally expensive.
Thus, larger datasets of chemically diverse molecules are commonly enumerated by using lower-fidelity approaches.
The GEOM dataset \cite{axelrod2022geom} employs XTB-CREST \cite{pracht2020automated} to cheaply enumerate and relax multiple low-energy conformers of large molecules.
This dataset can be split into two parts.
\textbf{GEOM-QM9} contains the same set of molecules as QM9, but enumerates a total of $1.82\times 10^6$ conformers by using XTB-CREST, for an average of roughly 14 conformers per molecule.
\textbf{GEOM-Drug} comprises $2.91 \times 10^5$ drug-like molecules identified across several sources, and enumerates a total of $3.12 \times 10^7$ conformers, for an average of $\sim$107 conformers per molecule.
For our isospectral evaluation, we consider only the lowest-energy conformer available per molecule in GEOM-Drug.
The GEOM dataset does not include dipole calculations, therefore in instances where a collision is deemed possible (based on $R,A,B,C$), we perform an XTB-GFN2 point calculation by using the available GEOM coordinates to determine and compare these principal axis oriented dipoles.

Finally, we draw all available geometries from \textbf{PubChem}, which totals over 110 million unique molecules, with one geometry per molecule and dipole moments generally unavailable \cite{kim2023pubchem}.
To the best of our knowledge, the PubChem dataset represents the largest set of aggregated molecule geometries currently available.
The fidelities of these geometries may vary widely, ranging from high-fidelity DFT approaches to simple force field relaxations.
We select from PubChem all molecules within a certain molecular weight range, which we obtain based on the hypothesized high structural diversity of molecules in the range, as we now describe.
\citet{luttschwager2013last} perform XTB simulations on successively longer alkane chains and find that at $C_{18}H_{38}$ (with a weight of $W_{LH} = 254$ Da), the alkane chain does not uphold a trans- orientation across all carbon-carbon bonds as the lowest-energy state, but  instead takes a cis- orientation on a middle carbon-carbon bond, forming a hairpin as the new lowest-energy state.
The transition of molecules with repeating subunits from a highly prolate configuration (trans- oriented bonds only) to a more spherical configuration (with some cis- oriented bonds) would suggest a high degree of structural diversity.
We therefore select a subset of PubChem structures in the range $W_{LH}\pm 10$ Da [i.e., $(244,264)$ Da], and only assess potential collisions on $R$, $A$, $B$, and $C$.
\autoref{tab:molecule_datasets} describes each of these molecule datasets by molecule/conformer count, level of theory, average molecular weight, and the availability of SMILES strings and available dipoles.

%\begin{table*}
\begin{table}
\caption{\label{tab:molecule_datasets}Properties of the seven molecule datasets considered in this work.}
%\centering
%\resizebox{\columnwidth}{!}{
%\scalebox{0.8}{
\begin{ruledtabular}
\begin{tabular}{crrccccc}
 &  &  &  & Average & Avail. & Avail. & \\
Dataset & Molecules & Conformers & Theory & Mol. Wt. & SMILES & Dipoles & Ref. \\
\hline
QM9 & $1.33 \times 10^5$ & $1.33 \times 10^5$ & B3LYP & 122.69 Da & Yes & Yes & \cite{ramakrishnan2014quantum}\\
%QM9 & $1.33 \times 10^5$ & $1.33 \times 10^5$ & B3LYP/6-31G(2df,p) & 122.69 Da & Yes & Yes & \cite{ramakrishnan2014quantum}\\
QM7x & 6950 & $4.03 \times 10^6$ & PBE0+MBD & 96.58 Da & No & Yes & \cite{hoja2021qm7}\\
%ANI-1x & 3114 & $4.96 \times 10^6$ & CCSD(T)*/CBS & 165.74 Da & No & Yes & \cite{smith2020ani}\\
GEOM-QM9 & $1.33 \times 10^5$ & $1.82 \times 10^6$ & XTB-GFN2 & 122.69 Da & Yes & Yes & \cite{axelrod2022geom}\\
GEOM-Drug & $2.91\times 10^5$ & $3.12 \times 10^7$ & XTB-GFN2 & 355.24 Da & Yes & Yes & \cite{axelrod2022geom}\\
GEOM-Drug (Top 1) & $2.91\times 10^5$ & $2.91\times 10^5$ & XTB-GFN2 & 355.24 Da & Yes & Yes & \cite{axelrod2022geom}\\
PubChem & $>$1.10 $\times 10^8$ & $>$1.10 $\times 10^8$ & Varied & 420.97 Da & Yes & No & \cite{kim2023pubchem}\\
PubChem ($W_{LH}\pm 10$) & $6.78 \times 10^6$ & $6.78 \times 10^6$ & Varied & 254.07 Da & Yes & No & \cite{kim2023pubchem}\\
\end{tabular}
\end{ruledtabular}
%} %Close resizebox
\end{table}
%\end{table*}

\section{Results}\label{sec:results}
We first consider the results of our constrained and unconstrained constructive geometries, then review the outcome of our isospectral funnel applied across molecular datasets.

\subsection{Constructive Isospectrality Approaches}

%\autoref{fig:isospec_knapsack} shows three examples of a near-isospectral collision (or twins) in the constrained environment (with 10, 20, and 30 point masses, respectively), as identified by using our iterative greedy approach and oriented according to the same principal rotation axes.
%The examples present a trend of how twins identified by a greedy additive approach become harder to generate as we increase the number of point masses.
First, we consider the inherent difficulty with generating isospectral pairs in the constrained environment.
This difficulty can be attributed to the combinatorial explosion of possible structures for the given set of point masses.
%Our analysis, further described in Appendix \ref{append:constrained_env} never uncovered an isospectral pair of any size (that was not isomorphic with respect to a translation, rotation, or reflection) in the constrained environment, and it is unclear whether such an isospectral pair could be constructed (either from structures in $\mathbb{R}^3$, or in $\mathbb{R}^n;\; n\geq 2$).
Our analysis, further described in Supplementary Information Section I, never uncovered an isospectral pair of any size (that was not isomorphic with respect to a translation, rotation, or reflection) in the constrained environment, and it is unclear whether such an isospectral pair could be constructed (either from structures in $\mathbb{R}^3$, or in $\mathbb{R}^n;\; n\geq 2$).

Compared to the constrained environment, generating twins to arbitrary numerical precision is straightforward in the unconstrained environment.
We also find that structures are not required to have the same number of point masses to identify isospectral collisions in an unconstrained environment, so long as the number of point masses exceeds three.
%Furthermore, an arbitrary number of distinct isospectral collisions can be achieved through various numerical approaches described in Appendix \ref{append:unconstrained_env}.
Furthermore, an arbitrary number of distinct isospectral collisions can be achieved through various numerical approaches described in Supplementary Information Section II.

\subsection{Molecule Analysis}
We begin by considering the distribution of molecules across datasets by using Ray's $\kappa$, then by exploring the isospectral collisions identified for each dataset.

\subsubsection{Dataset Summary}
\begin{figure}[b]
\includegraphics[width=0.65\textwidth]{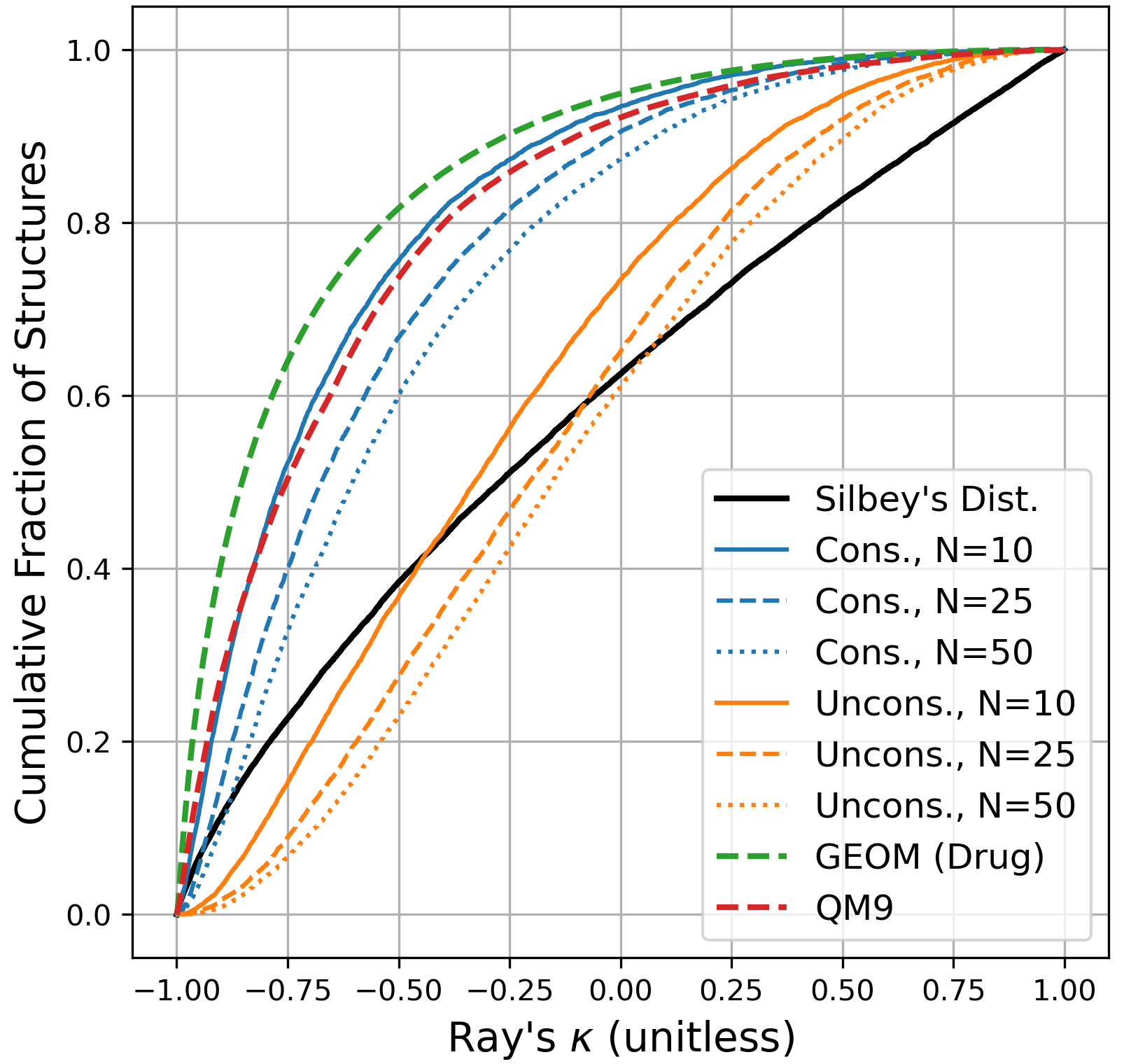}
\caption{\label{fig:rays_kappa_across_envs} Cumulative distribution of Ray's $\kappa$ across constrained (blue) and unconstrained (orange) environments with a varying number of point masses (10,25,50).
Silbey's cumulative probability distribution on $\kappa$ is also shown (black), along with the cumulative distribution of $\kappa$ within GEOM (Drug) and QM9 datasets (green and red, respectively).}
\end{figure}

\autoref{fig:rays_kappa_across_envs} shows the cumulative distribution of Ray's $\kappa$ across both constructed environments and molecule datasets, and also the theoretical cumulative distribution of $\kappa$ 
derived by \citet{silbey1988preponderance} from a closed-form expression based on an assumption of uniformly distributed moments of inertia among molecules.
The $\kappa$ distribution across all datasets shows that, in general, small to mid-sized organic molecules lean heavily towards a prolate rotor.
This prolate proclivity confirms earlier observations from \citet{silbey1988preponderance}, however the tendency for both random constrained structures and real molecules (QM9 and GEOM-Drug) to remain highly prolate is even more extreme than they first suggested.
Interestingly, we see that the distribution across $\kappa$ for real molecules more closely resembles the $\kappa$ distribution of the constrained environment than the unconstrained environment.

\begin{figure}[b]
\includegraphics[width=0.85\textwidth]{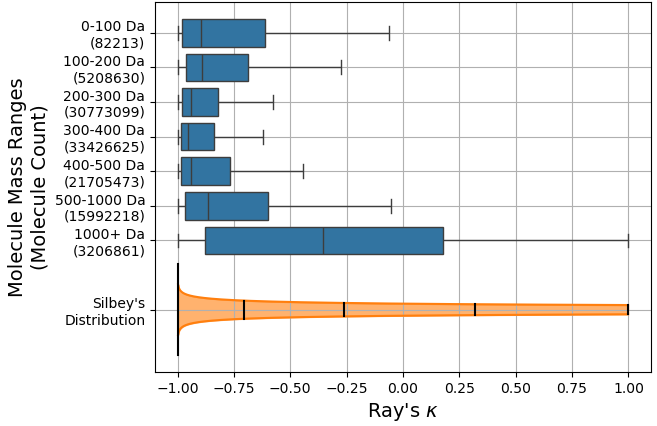}
\caption{\label{fig:pubchem_box_and_whisker} Box-and-whisker distribution of Ray's $\kappa$ for all molecules in the PubChem dataset, for binned masses in Daltons (Da).
The counts of PubChem molecules falling into each mass range is also listed.
Silbey and Kinsey's probability distribution on $\kappa$ is also shown \cite{silbey1988preponderance}, with black lines representing corresponding quartiles.}
\end{figure}

\autoref{fig:pubchem_box_and_whisker} shows a box-and-whisker plot of Ray's $\kappa$ across the set of PubChem molecules, separated by ranges in molecular weight.
Silbey's distribution is also shown with labelled quartiles.
%The distribution of PubChem molecular weights within these ranges is available in Appendix~\ref{append:pubchem_molecules}.
The distribution of PubChem molecular weights within these ranges is available in Supplementary Information Section III.
The distribution of structures between 0 and 100 Da is almost entirely prolate, but structures become even more concentrated at a prolate extreme at higher molecular weights.
We see an inflection in the 300 to 400 Da range (with average $\kappa \approx -0.92$), which is the most prolate range, after which molecules become more asymmetric (and even oblate).
The set of structures with molecular weight $>$1000 Da contains many oblate structures, and has an average $\kappa \approx -0.32$.

%In Appendix \ref{append:qm9_multifi}, we briefly compare the distribution of rotational constants and $\kappa$ between the lowest-energy conformers GEOM-QM9 dataset\cite{axelrod2022geom} and those in the high-fidelity DFT-optimized QM9 dataset\cite{ramakrishnan2014quantum}.
In Supplementary Information Section V, we briefly compare the distribution of rotational constants and $\kappa$ between the lowest-energy conformers GEOM-QM9 dataset\cite{axelrod2022geom} and those in the high-fidelity DFT-optimized QM9 dataset\cite{ramakrishnan2014quantum}.
We see a shift towards a more prolate extreme with high-fidelity DFT geometries, although it is unclear whether this trend continues at even higher-fidelity DFT methods.

\subsubsection{Molecule Isospectral Analysis}

\begin{figure}[b]
\includegraphics[width=0.75\textwidth]{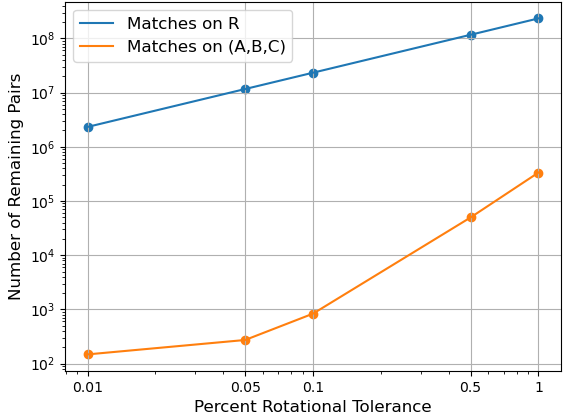}
\caption{\label{fig:qm9_tolerance_narrowing} The remaining number of twin pairs in QM9 based on the overall rotational inertia $R$ as well as rotational constants $(A,B,C)$.
The number of remaining twin pairs becomes progressively smaller as tolerance decreases from 1\% to 0.01\%.}
\end{figure}

\autoref{tab:molecule_isospec} shows the number of possible twin pairs (within pre-defined tolerances) across molecule datasets along each step of the funnelling process.
%First, we see that collisions using only the overall rotational constant $R$ represent $\sim$1/1000 of all possible conformer pairs.
For datasets containing only a single conformer per distinct molecule (QM9, GEOM-Drug, and PubChem), we find that funnelling on $R$ results in about 1/100 remaining pairs.
For QM9 and GEOM-Drug, we see that collisions on rotational constants are rare (roughly 1/20000).
When many conformers of the same molecule are considered (as with QM7x and GEOM-QM9), many more matches on rotational constants occur.
These collisions among structures can come from conformers of either the same or different molecules.
Comparing collisions for dipole moment ratios narrows the set of possible pairs once more: for datasets with a single conformer per molecule, the pair reduction is roughly a further three orders of magnitude.
\autoref{fig:qm9_tolerance_narrowing} shows the effect of changing the tolerance on $R$ and $(A,B,C)$ over the range 0.01\% to 1\%.
If rotational and dipole constants could be obtained via simulation at an even higher level of accuracy, the number of remaining twin pairs could decrease yet further.

\begin{table*}
\caption{\label{tab:molecule_isospec}Remaining twins from isospectral funnelling of conformers across various datasets.}
\begin{ruledtabular}
\begin{tabular}{ccccccc}
Dataset & \begin{tabular}[c]{@{}l@{}}Considered\\Conf. Pairs\end{tabular} & \begin{tabular}[c]{@{}l@{}}Overall Rot.\\ Collision\end{tabular} & \begin{tabular}[c]{@{}l@{}}Rot. Const.\\ Collision\end{tabular} & \begin{tabular}[c]{@{}l@{}}Dipole\\ Collision\end{tabular} & \begin{tabular}[c]{@{}l@{}}Stereo.\\ Collision\end{tabular} & \begin{tabular}[c]{@{}l@{}}Formula\\ Collision\end{tabular} \\
\hline
QM9 & $8.45 \times 10^9$ & $2.34 \times 10^8$ & $3.35 \times 10^5$ & 356 & 295 & 36 \\
QM7x & $8.11\times 10^{12}$ & $1.91 \times 10^{11}$ & $3.46 \times 10^8$ & $1.38 \times 10^6$ & - & $1.89 \times 10^5$ \\
GEOM-QM9 (All) & $1.66 \times 10^{12}$ & $4.35 \times 10^8$ & $5.85 \times 10^5$ & $5.98 \times 10^3$ & 349 & 27 \\
GEOM-Drug (Top 1) & $4.23\times 10^{10}$ & $4.90 \times 10^8$ & $4.25 \times 10^5$ & 941 & 621 & 462 \\
PubChem ($W_{LH}\pm 10$) & $2.30 \times 10^{13}$ & $3.23 \times 10^{11}$ & $6.36 \times 10^9$ & - & - & - \\
\end{tabular}
\end{ruledtabular}
\end{table*}

\begin{figure}[b]
\includegraphics[width=0.49\textwidth]{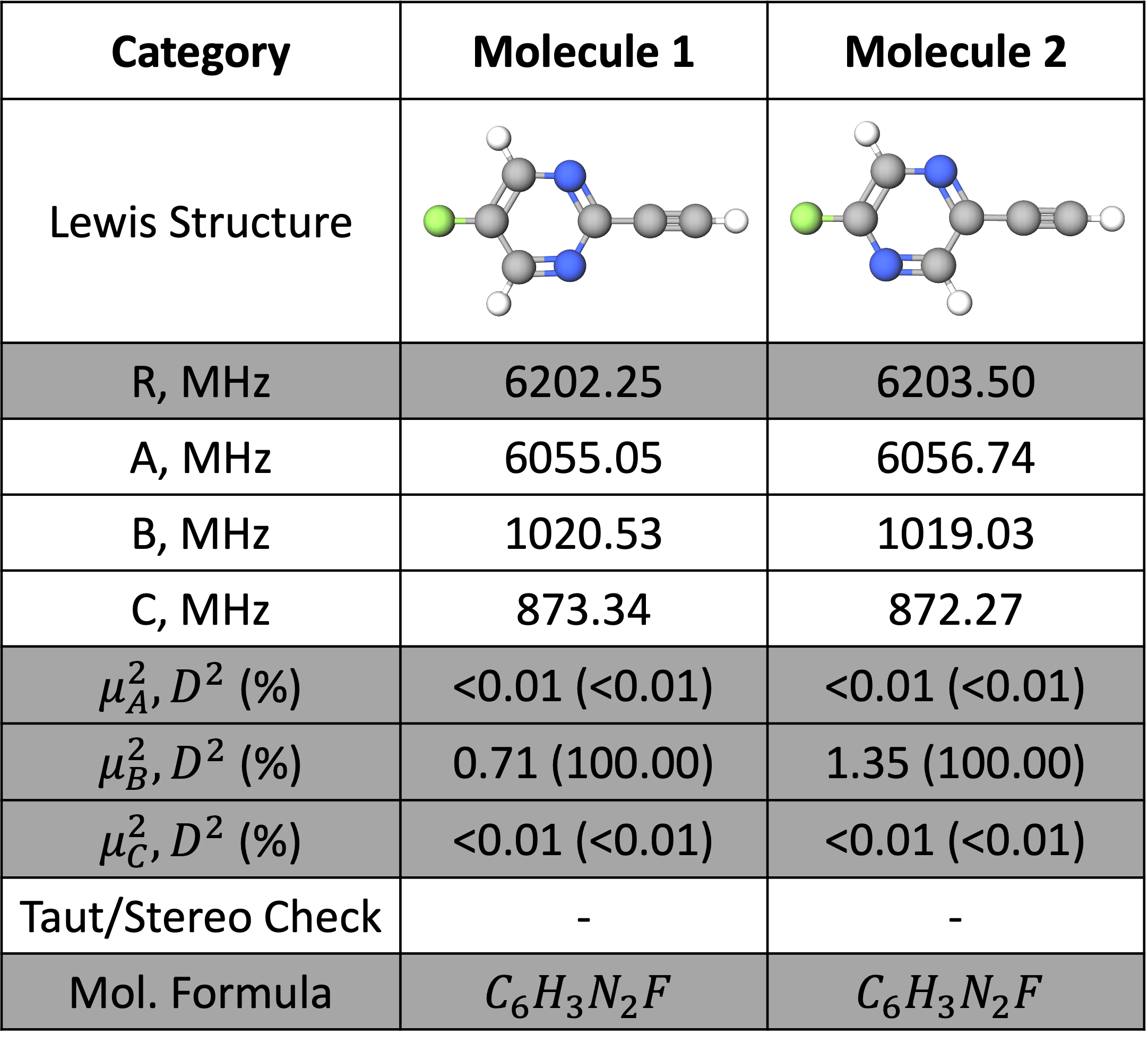}
\includegraphics[width=0.49\textwidth]{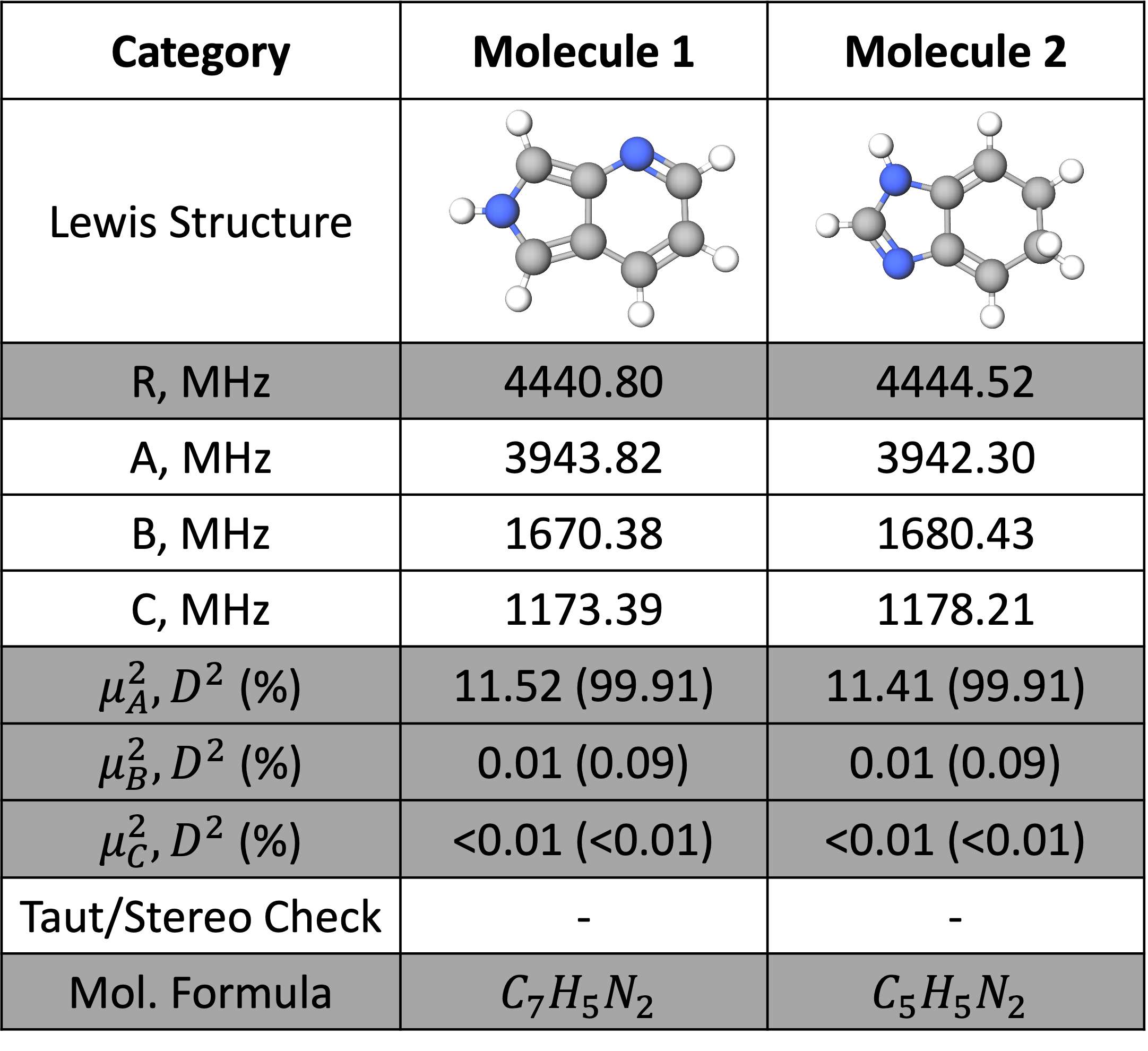}
\caption{\label{fig:QM9_ex_struct_iso}
Two pairs of low-energy molecule conformers in the QM9 dataset.
In each, Molecule 1 and Molecule 2 are
both twins and structural isomers.
}
\end{figure}

\begin{figure}[b]
\includegraphics[width=0.49\textwidth]{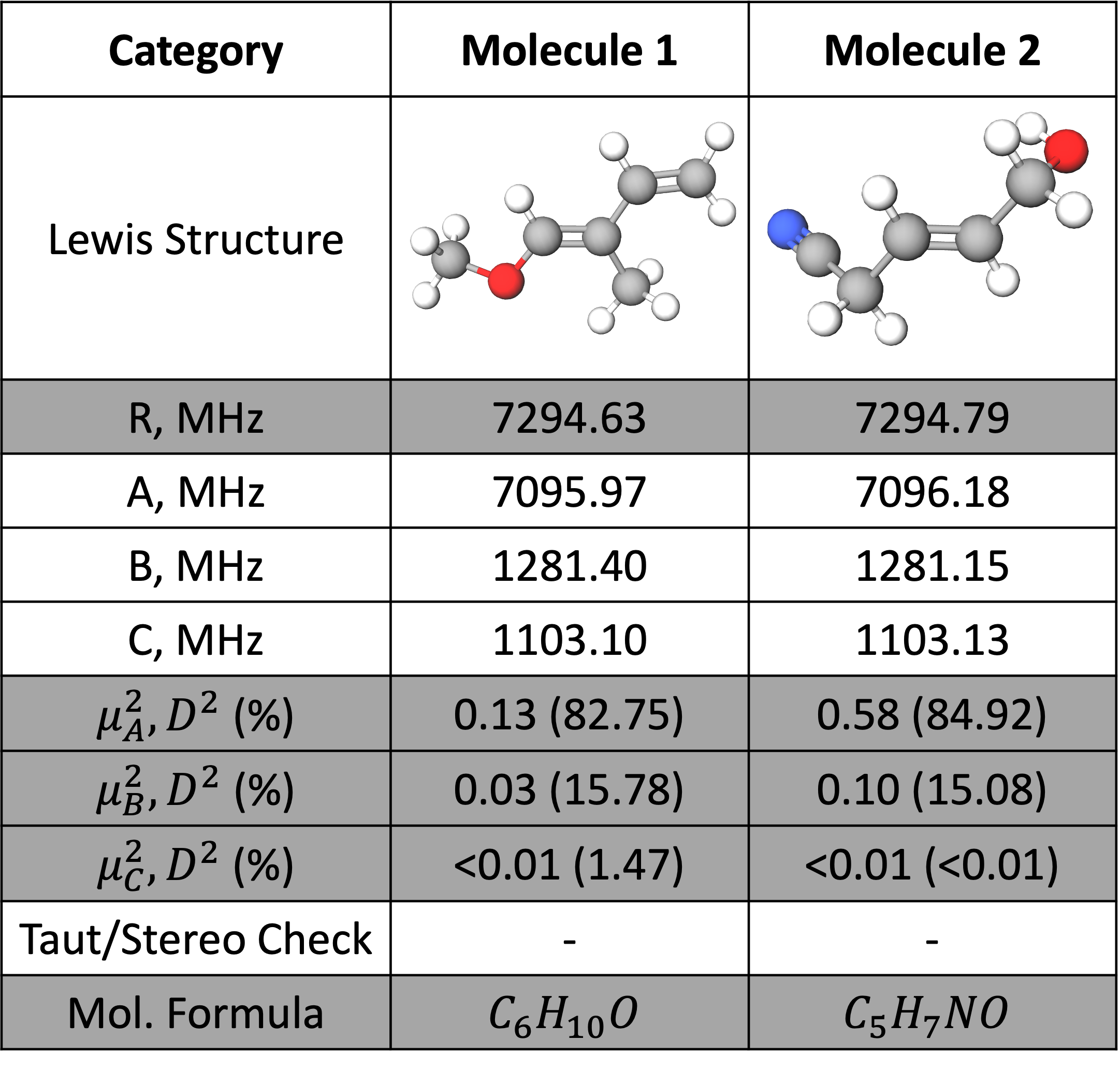}
\includegraphics[width=0.49\textwidth]{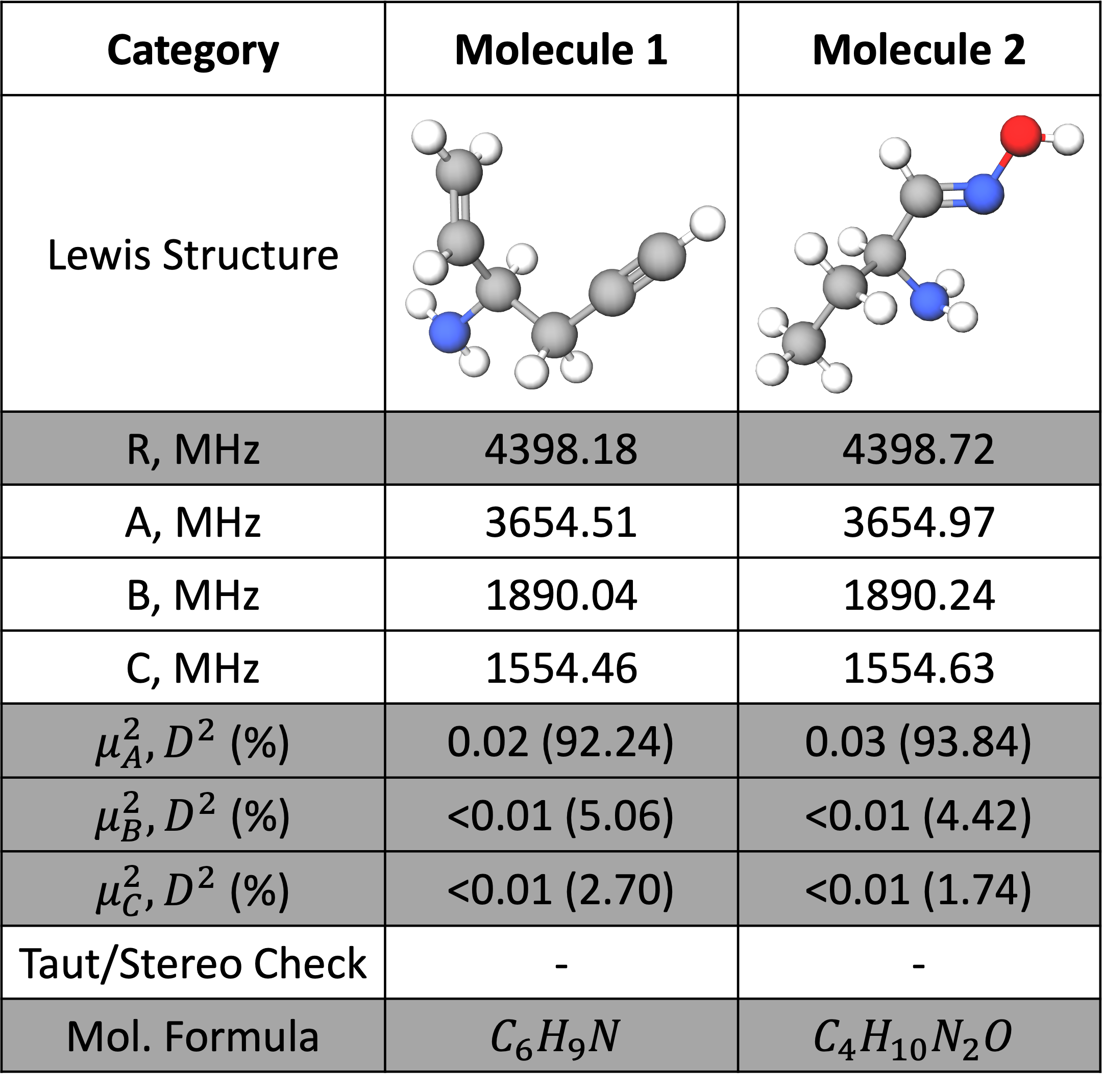}
\caption{\label{fig:QM7x_examples} 
Two pairs of molecule conformers in the QM7x dataset that are twins.
%In each case, the right molecule differs from the left in the substitution of C for N.
}
\end{figure}

%We now identify a number of remaining twins across each dataset to characterize the sorts of molecules which remain after funnelling.
We now consider a number of twins across each dataset to characterize the sorts of molecules which remain after funnelling.
\autoref{fig:QM9_ex_struct_iso} shows two examples of structural isomers from QM9 with very similar rotational constants and a single strong dipole component ($\mu_B$).
Note that $^2 H$ or $^{15} N$ isotopic substitution on each species would produce distinct rotameters which, once labelled, would remove structural ambiguity.
\autoref{fig:QM9_ex_struct_iso} shows examples of twins which are not structural isomers in QM9, and could in practice be distinguished by using mass spectrometry.
If such a measurement cannot be taken, these molecules are also more conformationally flexible and could therefore be distinguished by the presence of other, distinct low-energy conformers.
For a case where both species could be present in the same sample, a nutation experiment could provide distinguishing information about the relative dipole intensities of both species (without performing a Stark measurement).
\autoref{fig:QM7x_examples} shows examples of twin conformers in QM7x, which occur alongside a number of other conformers for both species.
In practice these spectra are distinct when multiple conformers are present, with the exception of this pair of twin conformers.

\begin{figure}[b]
\includegraphics[width=0.49\textwidth]{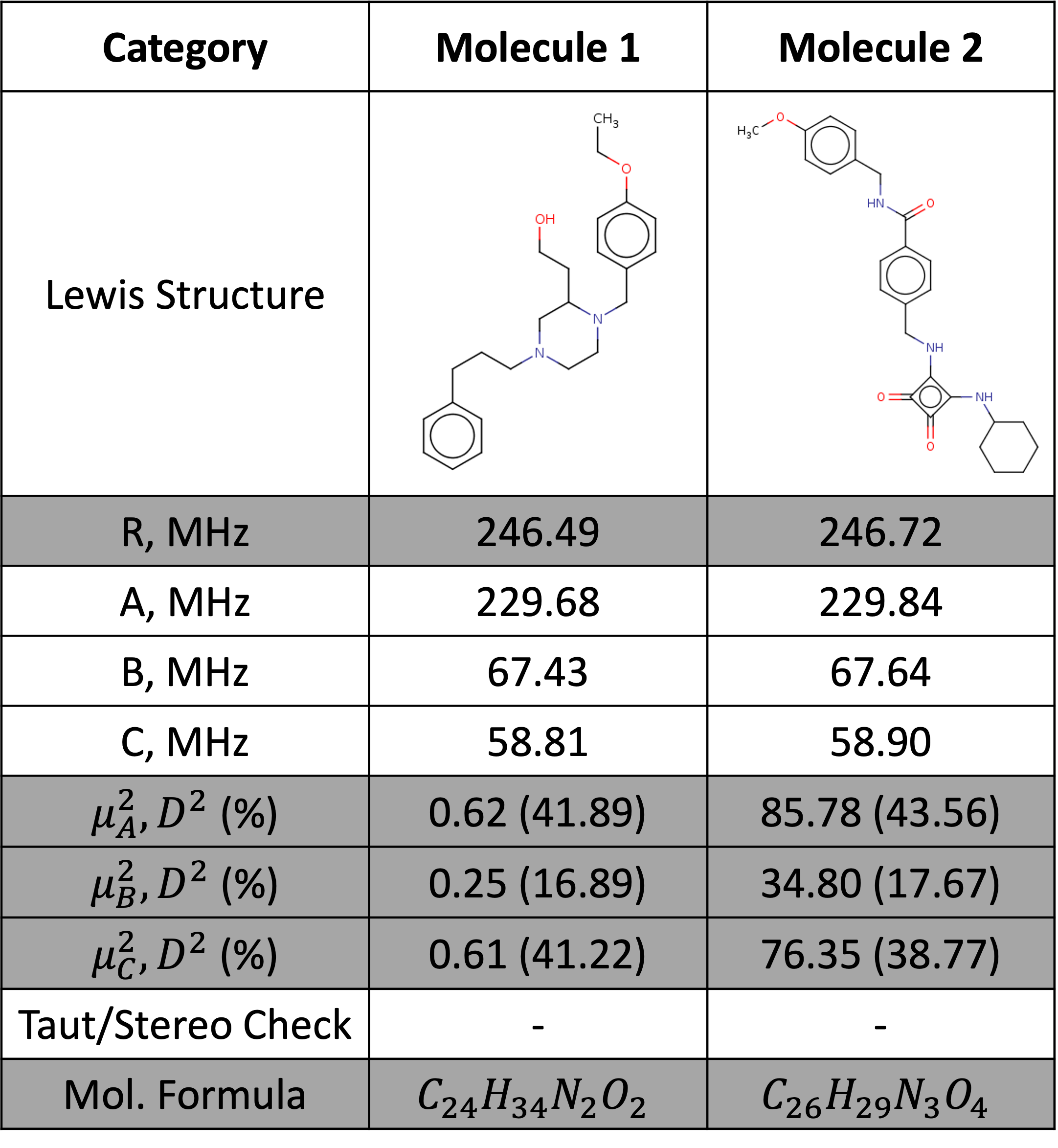}
\includegraphics[width=0.49\textwidth]{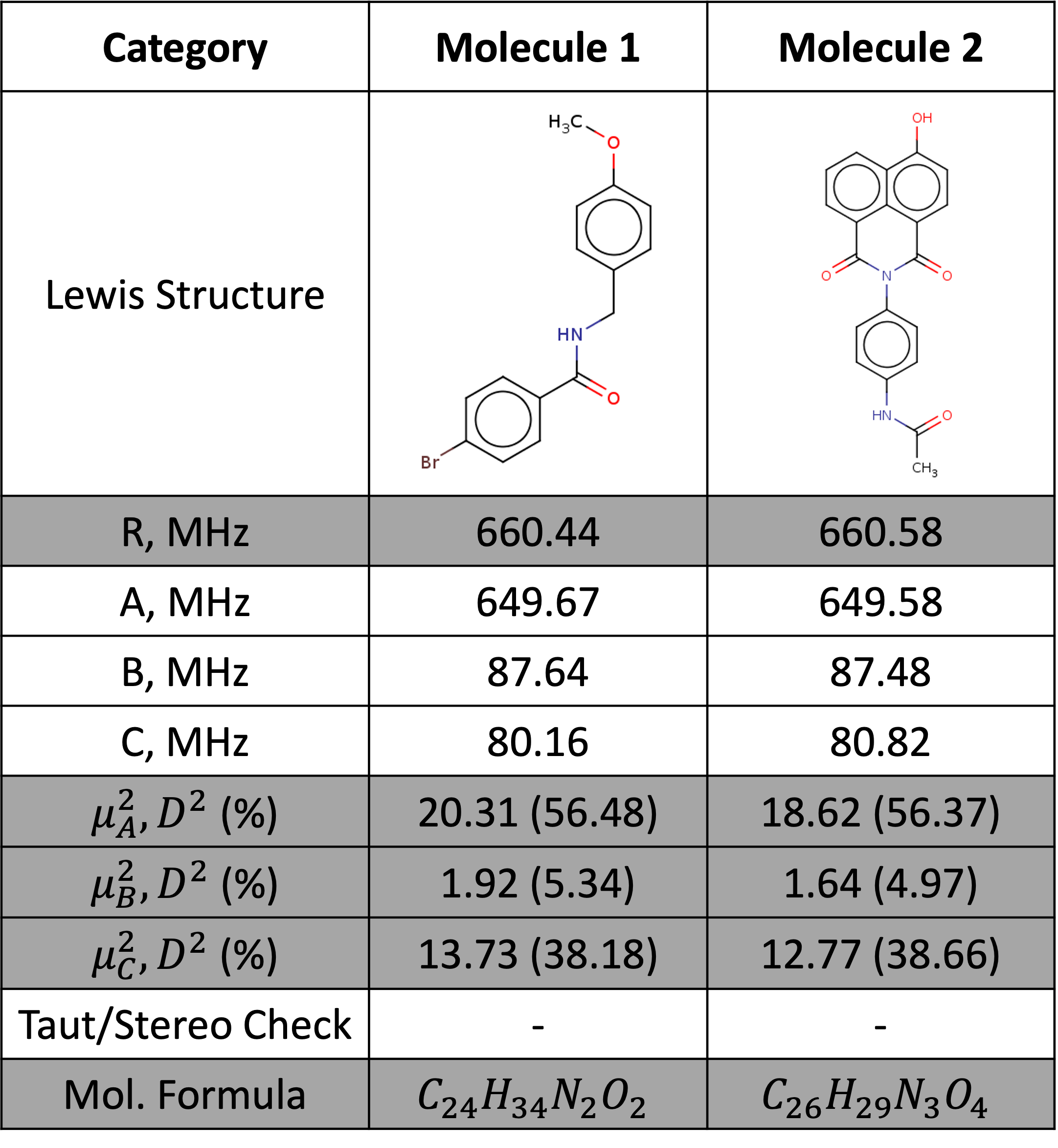}
\caption{\label{fig:GEOM_drug_examples} Two pairs of molecule conformers in the GEOM-Drug dataset that are twins.}
\end{figure}

Finally, \autoref{fig:GEOM_drug_examples} shows examples of molecules from the GEOM-Drug dataset.
It is unclear whether these molecules could be observed via rotational spectroscopy, even when using ablative techniques.
In practice (and depending on conformational temperature), both molecule conformations are present at different abundances among a variety of other conformational modes.

\begin{figure}[b]
\includegraphics[width=0.49\textwidth]{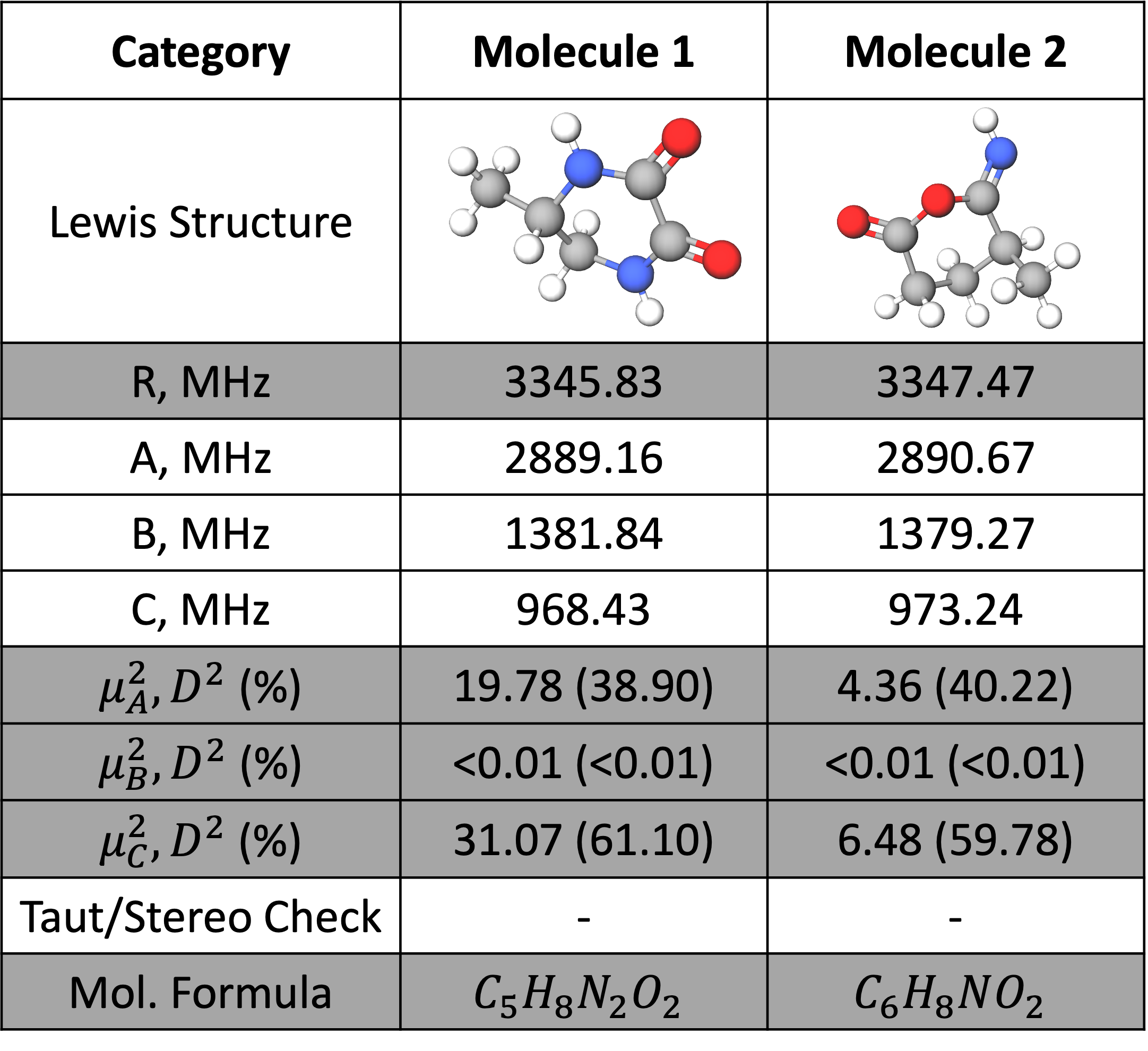}
\includegraphics[width=0.49\textwidth]{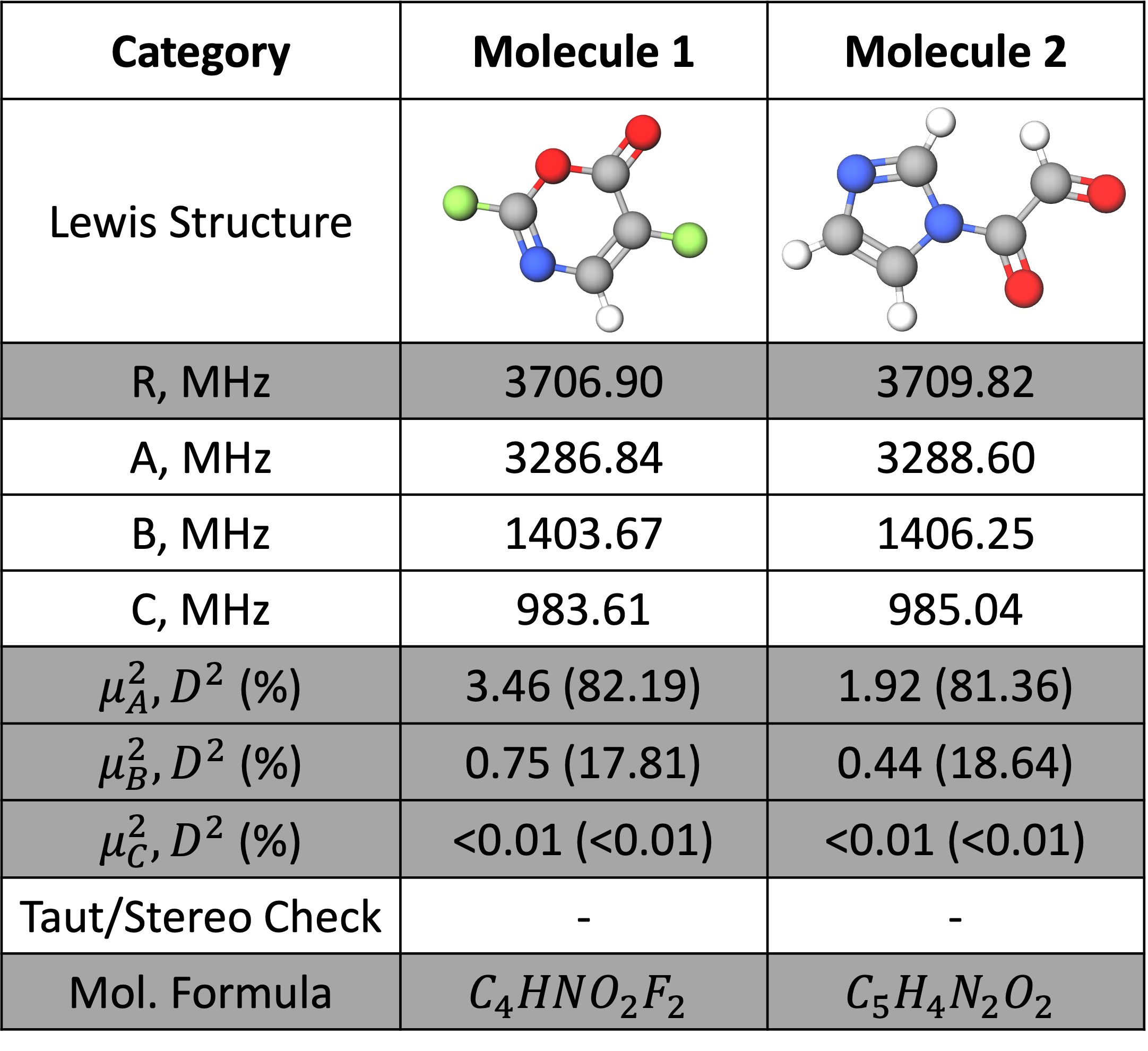}
\includegraphics[width=0.49\textwidth]{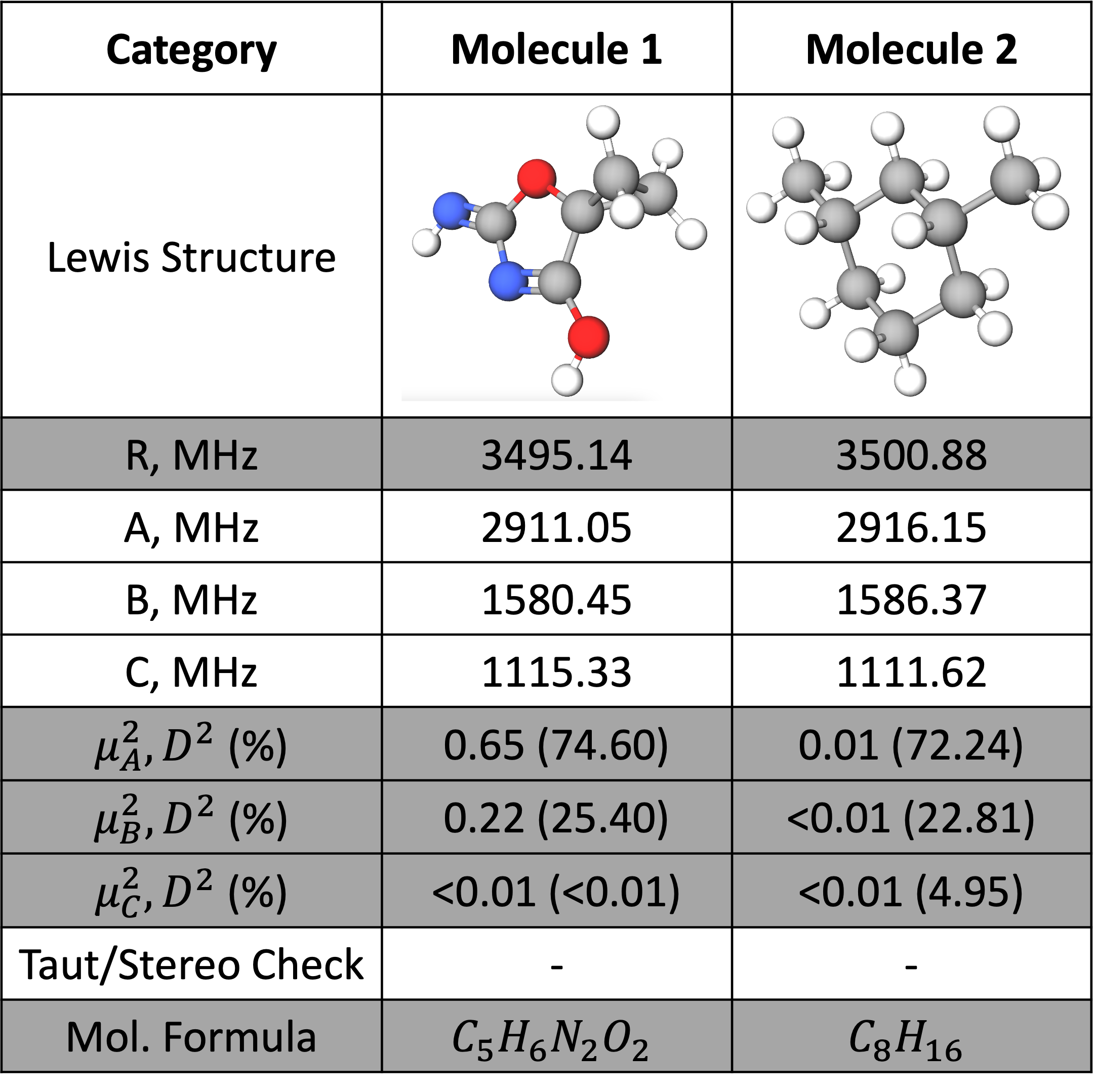}
\includegraphics[width=0.49\textwidth]{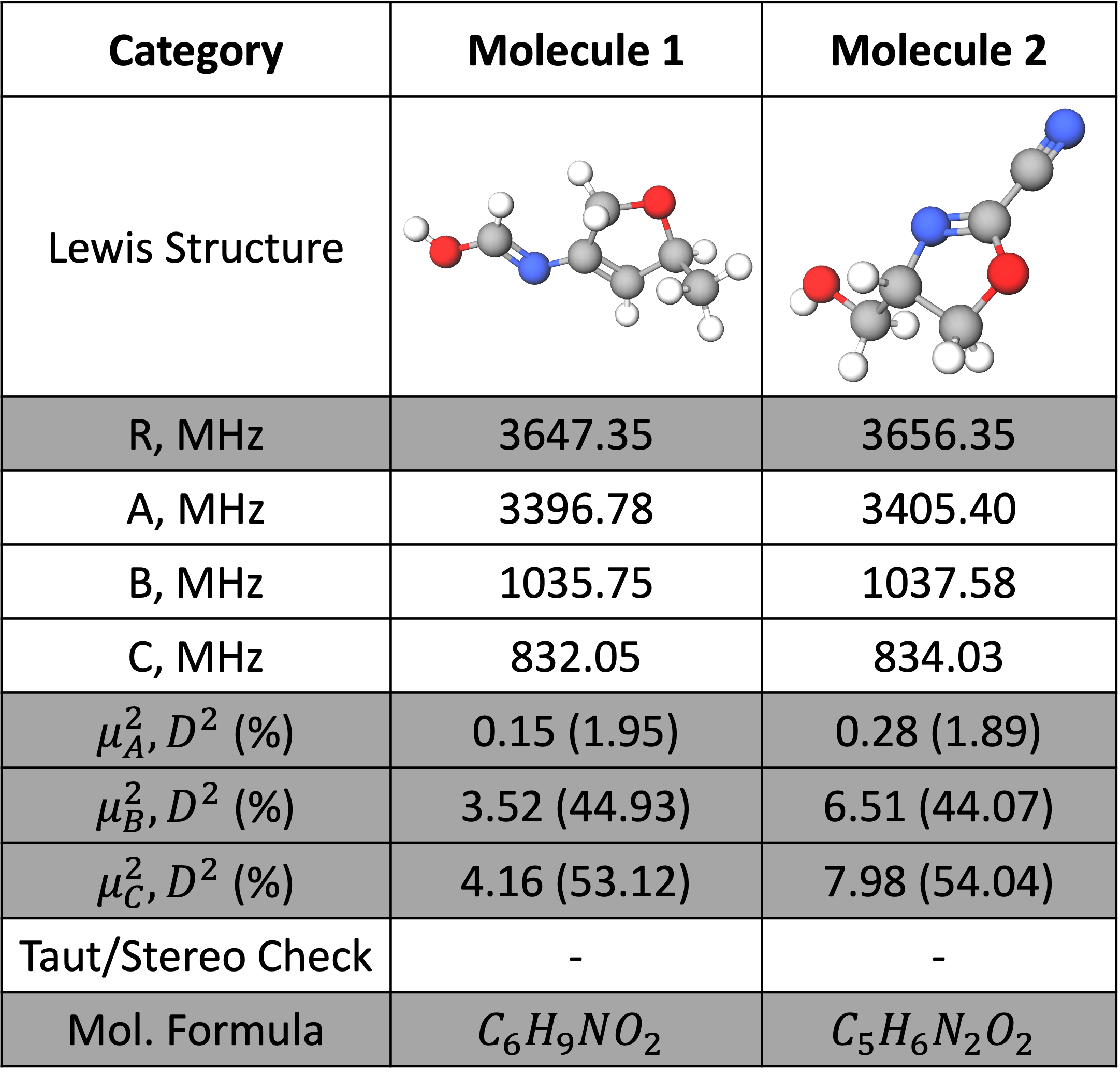}
\caption{\label{fig:QM9_ex_distinct} 
Four pairs of low-energy molecule conformers in the QM9 dataset.
In each, Molecule 1 and Molecule 2 are twins but not structural isomers.}
\end{figure}

\section{Discussion}\label{sec:discussion}
We assert that, with respect to structural diversity, the space of possible molecule conformers lies somewhere between the two constructed extremes we have presented.
Of course, atoms in a conformer are not constrained to a single discrete mass, bonded atom pairs may be of varying lengths, and bond angles are not limited to $90^\circ$.
Conversely, atoms in a conformer are not fully unconstrained with regard to masses, bond lengths, and bond angles.
It is unclear whether the properties of the space of conformers more closely resemble the constructed constrained or the unconstrained environment.
However, based on the distribution of Ray's $\kappa$ in \autoref{fig:rays_kappa_across_envs}, the space of molecule conformers seems to resemble the constrained environment in terms of $\kappa$.

%Forward: structures to moments of inertia [we say this is well-posed]
%Inverse: constants --> structures
The forward mapping of structures in the constrained environment to moments of inertia appears well-posed, since no isospectral collisions were observed.
By contrast, the function mapping structures in the unconstrained environment to moments of inertia is ill-posed, with an infinite number of distinct structures capable of satisfying the same set of moments of inertia.
If the space of molecules resembles the constrained environment, as suggested by Figure \ref{fig:rays_kappa_across_envs}, this may imply that no indistinguishable collision exists.

From our analysis of conformer datasets, we find that many pairs of conformers have rotational constants within our specified tolerances.
However, far fewer twin pairs are identified that have both rotational and dipole constants within our specified tolerance.
The examples in \autoref{fig:QM9_ex_struct_iso} are among the closest matches between molecules in QM9, and both pairs of similar structures offer little conformational flexibility.
Supposing these molecules were present in a mixed sample, in practice they could likely be distinguished from one another quite readily, although their respective identities may remain ambiguous. 
As previously mentioned, it is common for rotational spectrometers to assess peak frequencies with $\sim$10~kHz accuracy, which corresponds to the relative uncertainty in measured rotational constants of $\Delta\chi/\chi \sim 10^{-4} - 10^{-6}, \chi \in [A, B, C]$. At the same time, the DFT-calculated uncertainty is $\Delta\chi/\chi \sim 10^{-2}$.
In other words, two molecules may be twins because they are experimentally distinguishable, but %uncertainty 
discrepancy between measurement and simulation prevents us from accurately identifying which species aligns with which peaks in an experimental spectrum.
%As previously mentioned, the high signal-to-noise ratio on most rotational spectrometers allows for peak frequencies to be accurately assessed within 10 kHz, so differences could probably be easily observed between this pair of molecules.%

%so differences could probably be clearly observed between this pair of molecules.
%Due to the error between simulation and experiment, it remains unclear whether twins will always be distinguishable.
%even prior to assessment from quadrupole coupling constants.
The examples from QM7x in \autoref{fig:QM7x_examples} show another pair of twin conformers. However, in this case the comparison is between two conformers of different molecules where each molecule exhibits many possible conformers.
In practice, these molecules could be distinguished from one another by using other conformers.
In all cases presented, isotopomeric species can play a part in distinguishing these molecules as well.
Thus, even in instances where near-isospectral matches between molecules or conformers appear in the structure datasets we evaluated, the pairs can be readily distinguished by using some combination of high signal-to-noise ratios, nutation, conformational flexibility, isotopomeric substitution, or molecular formula information from mass spectrometry.
Instances where this additional level of analysis is required to distinguish between species appear to be rare in practice.

%Considering the rate of potential twin pairs with increasing molecule size, it is often regarded as easier to distinguish between larger molecules because their smaller rotational constants will lead to a greater number of spaced transition lines within a band region.
%This effect provides more discriminating information in a fixed bandwidth, even as the space of distinct molecules and conformers explodes combinatorially.
%We can see this effect in practice when comparing funnelling between QM9, with average molecular weight of 122.69 Da, and GEOM-Drug (Top 1), with average molecular weight of 355.24 Da.
We also consider the number of remaining twins for molecules of varying masses.
We compare QM9, with average molecular weight of 122.69 Da, and GEOM-Drug (Top 1), with average molecular weight of 355.24 Da, to see how changes in mass influences the number of remaining twins.
Even though the set of conformer pairs in GEOM-Drug (Top 1) was roughly five times the size of the set of conformer pairs in QM9, the number of remaining twin pairs after funnelling by $R$ and $(A,B,C)$ is a smaller fraction of total possible conformer pairs.
Since rotational spectroscopy is a gas-phase technique, compounds with low volatility or high boiling points may not be measurable.
While this may imply a ceiling on the size of molecules that can be analyzed by using rotational spectroscopy, a number of ablation techniques have been devised to coax large aromatic molecules into the gas phase.\cite{hensel1993microwave,lesarri2003laser}
Molecular complexation can also be measured for a number of species by using rotational spectroscopy at specific conditions, with correspondingly low rotational constants permitting many molecule complex conformers to be uniquely identified \cite{perez2012structures}.

\section{Conclusion}\label{sec:conclusion}
Advancement of broadband rotational spectroscopy to the realm of analytical chemistry relies on its capacity to discern robustly between distinct molecules.
This discerning power can be mathematically expressed through the well-posedness of the inverse problem that maps spectra to molecular structures.
While we know that a molecular structure that gives rise to a spectrum exists, in this work we have explored whether the structure that produces such a spectrum is unique, which would make the inverse problem well-posed.

First, we construct constrained and unconstrained environments and assess how isospectral collisions\textemdash the instances of different molecular structures having an indistinguishable set of rotational constants\textemdash can be identified.
We find that (contrived) \textit{constrained} environments produce structures more similar to real molecules (according to Ray's $\kappa$), and do not yield any isospectral collisions.
In contrast, the spatially \textit{unconstrained} assembly of point masses readily leads to collisions with arbitrary numerical precision.

Second, we search several large datasets of calculated molecular geometries for potential isospectral collisions by using a funnelling approach.
The number of collisions falls rapidly as the number of parameters to be matched (such as rotational constants and dipole moment projections) increases, and as the allowed uncertainties in these parameters are tightened.
We found instances of molecule twins, which have predicted rotational and dipole constants close enough that a standard molecular simulation would not be able to discern which spectrum corresponds to which of the two molecules, even if their spectra were measured to be distinct.
It is possible that with higher accuracy calculations of zero-point-averaged molecular structures or with additional experiments (such as isotopic substitution or nutation) these collisions could be resolved.
Therefore, we conclude that for molecules in the present datasets, the mapping from spectra to structures may be well-posed in principle, but is ill-posed at the current level of accuracy offered by reasonably fast calculations of structures.
Although we only identify twin pairs in this work, it remains unclear whether any pairs of molecules with experimentally indistinguishable rotational spectra will be identified in practice.

\section{Future Work}
Several other possibilities could be considered when assessing isospectrality constraints.
First, instead of comparing individual conformer pairs against one another, one could perform a comparison across sets of conformers associated with separate molecules.
A collision across molecules with multiple conformers would be far less likely than a collision between two conformers, and Boltzmann-weighted conformer abundances (assuming a thermodynamic equilibrium distribution of conformers) would also need to be considered alongside dipole magnitudes.
Next, rather than comparing only the constants pertaining to specific structures, another framing of the isospectrality question might compare generated spectra by using an optimal transport distance, Hamming, or Minkowski $P\neq 2$ metric.
Finally, assessing larger datasets of geometries and conformers might still uncover twin conformers that are irreconcilable from either a theoretical or an experimental perspective.
Twins may also be identifiable among distinct complexes of molecules, which may exhibit rotational symmetries that single molecules cannot emulate.
In this work we limit our consideration to pure rotational spectroscopy, but the similar questions are equally valid for other spectroscopic modes.

\section*{Data Availability}
The code associated with the findings in this paper are available via GitHub: [].
The data that support the findings in this paper are available from the corresponding authors upon reasonable request.

\section*{Author Declarations}
K.\,P. holds patents US11380422B2 and US11594304B2 that are related to this work.

\begin{acknowledgments}
The authors thank Logan Ward for his thoughts, comments, and ideas on isospectrality in spectroscopy.
This material is based on work supported by the U.S. Department of Energy, Office of Science, Office of Basic Energy Sciences, Division of Chemical Sciences, Geosciences, and Biosciences under Contract No. DE-AC02-06CH11357.
\end{acknowledgments}

%END MULTI-LINE COMMENT

\bibliography{aipsamp}% Produces the bibliography via BibTeX.

\end{document}

\preprint{AIP/123-QED}

%\title[Sample title]{Sample Title:\\with Forced Linebreak\footnote{Error!}}% Force line breaks with \\
%\thanks{Footnote to title of article.}
%\title{Can One Hear the Shape of a Molecular Rotor?\\The Rotational Spectroscopy Inverse Problem
%}
\title{
%Isospectrality in rotational spectroscopy: does a rotational spectrum uniquely identify a molecule?
Supplementary Information for "Twins in rotational spectroscopy: Does a rotational spectrum uniquely identify a molecule?"
}

\author{Marcus Schwarting}
\affiliation{Department of Computer Science, University of Chicago, Chicago, IL 60637, USA}%Lines break automatically or can be forced with \\

\author{Nathan~A.~Seifert}%
\affiliation{ 
Department of Chemistry and Chemical \& Biomedical Engineering, University of New Haven, West Haven, CT 06516, USA
}%

\author{Michael~J.~Davis}
\affiliation{%
Chemical Sciences and Engineering Division, Argonne National Laboratory, Lemont, IL 60439, USA}%

\author{Ben Blaiszik}
\affiliation{Data Science and Learning Division, Argonne National Laboratory, Lemont, IL 60439, USA}

\author{Ian Foster}
\affiliation{Department of Computer Science, University of Chicago, Chicago, IL 60637, USA}
\affiliation{Data Science and Learning Division, Argonne National Laboratory, Lemont, IL 60439, USA}

\author{Kirill Prozument}
\affiliation{Chemical Sciences and Engineering Division, Argonne National Laboratory, Lemont, IL 60439, USA}

\date{\today}% It is always \today, today,

\maketitle

%\begin{quotation}
%The ``lead paragraph'' is encapsulated with the \LaTeX\ 
%\verb+quotation+ environment and is formatted as a single paragraph before the first section heading. 
%(The \verb+quotation+ environment reverts to its usual meaning after the first sectioning command.) 
%Note that numbered references are allowed in the lead paragraph.
%
%The lead paragraph will only be found in an article being prepared for the journal \textit{Chaos}.
%\end{quotation}

%TODO: INCLUDE A BRIEF EXPLANATION OF WHAT IS GOING ON HERE...? Summarize the sections?

\section{The Constructed Constrained Environment}\label{append:constrained_env}
In this section, we go into greater depth on the structures created in the constrained environment.
Given an arbitrary existing structure $\mathcal{S}$, candidate positions for adding another point mass are compiled according to
$$\mathcal{P_S} = \{p=(x_i\pm1,y_i,z_i),(x_i,y_i\pm1,z_i),(x_i,y_i,z_i\pm1) \vert (x_i,y_i,z_i)\in \mathcal{S}; p \notin \mathcal{S} \}.$$
In the case where a target set of rotational inertias$\boldsymbol{I_T}=(I_A,I_B,I_C)$ is specified, a position may be selected to minimize the squared L2 loss function
$$\ell(\mathcal{S} \vert \boldsymbol{I_T}) =  \vert \vert \theta_T(\mathcal{S}), \boldsymbol{I_T} \vert \vert_2^2$$
where $\theta_T$ is a function which takes a structure and returns the corresponding ordered inertial parameters as described above.
%, and $\vert \vert \cdot, \cdot \vert \vert_2$ is the Euclidean norm.
If a random structure is desired, a new position is selected from $\mathcal{P_S}$ uniformly at random.

This combinatorial optimization problem lends itself naturally to a greedy packing strategy similar to what might be employed in an unconstrained knapsack problem \cite{akccay2007greedy}.
%Furthermore, as this toy problem is framed, it can be shown that it is an NP-optimization problem \cite{cai1997fixed}.
Furthermore, this problem is framed as an NP-optimization problem \cite{cai1997fixed}.
We speculate that it may be possible to identify isospectral collisions by using a spectral graph theory approach, akin to those used with HMO isospectrality described above \cite{chung1997spectral}.

Algorithm \ref{algo:greedy_inertia} implements a greedy process (mirroring greedy packing) for adding point masses to a structure to approximate target inertias $\boldsymbol{I_T}$ as closely as possible.
%While the process appears fully deterministic, we find that c
We find in practice that calculating $\ell(\cdot \vert \boldsymbol{I_T})$ often results in ties, in which case the next added point mass $p_j^*$ is selected randomly from among these ties, adding a level of stochasticity to an otherwise deterministic process.
We therefore use $N$ random restarts to increase the chances that a structure more closely approximates the target inertias.

\begin{algorithm}
\caption{Greedy algorithm with $N$ restarts to identify a structure $\mathcal{S}^*$ with moments of inertia approaching $\boldsymbol{I_T}$.}\label{algo:greedy_inertia}
\textbf{Initialize:} Target moments of inertia $\boldsymbol{I_T}=(I_A,I_B,I_C)$, $N$ random restarts, $n$ point masses.\\
\For{$i=0,...,N$}
{
    $\mathcal{S}_i = \{(0,0,0) \}$\\
    \For{$j=1,....(n-1)$}
    {
        %$p^*_j = \ell(\mathcal{S}_i \vert \boldsymbol{I_T})$\\
        Uniformly select $p^*_j \in \underset{p \in \mathcal{P_S}}{\mathrm{argmin}} \bigr\{ \ell(\mathcal{S}_i \cup \{p\} \vert \boldsymbol{I_T}) \bigr\}$\\
        $\mathcal{S}_i \gets \mathcal{S}_i \cup \{p^*_j \}$
    }
}
\Return{$\mathcal{S}^* = \underset{\mathcal{S}_i \in \{\mathcal{S}_0,...,\mathcal{S}_N \}}{\mathrm{argmin}}  \bigr\{ \vert \vert \theta_T(\mathcal{S}_i),  \boldsymbol{I_T} \vert \vert_2^2 \bigr\}$}
\vspace{1ex}
\end{algorithm}

\section{The Constructed Unconstrained Environment}\label{append:unconstrained_env}
Here we go into further details on the process by which we identify isospectral collisions in the unconstrained environment.
First, we derive the expression which led to identifying the isospectral collision with $\mathcal{S}_1 = \{ (0,1,0),(1,0,0),(0,-1,0)\}$ with corresponding masses $(1,2,1)$ and $\mathcal{S}_2 = \{ (0,1,0),(\sqrt{3},0,0),(0,-1,0) \}$ with corresponding masses $(1,1,1)$.
Note that the first and third data points are identical in the two structures.
Suppose we only allow the x-coordinate and mass of the second point to vary.
That is, suppose we have the set of structures $\mathcal{S}_k = \{ (0,1,0),(x_k,0,0),(0,-1,0) \}$ with masses $(1,m_k,1)$.
We can show that $\bar{x} = \frac{xm}{(m+2)};\; \bar{y}=0;\; \bar{z}=0$.
From here, we can show that
$$I_{x,x}=2;\;I_{y,y}=\frac{2x_k^2m_k^2}{m_k+2};\;I_{z,z}=I_{y,y}+2;\;I_{x,y}=I_{x,z}=I_{y,z}=0.$$
Since the off-diagonal terms are all zero, and since one can control diagonal elements by changing only the $I_{y,y}$ term, we can fix $I_{y,y}$ at any arbitrary positive value and derive infinitely many pairs of $(x_k,m_k)$ that satisfy isospectral constraints.
We may even rearrange our formulation in terms of $m_k$ or in terms of $x_k$, respectively:
$$m_k = \frac{I_{y,y}+\sqrt{I_{y,y}^2 + 16x_k^2I_{y,y}}}{4x_k^2};\;
x_k = \sqrt{I_{y,y}\biggr( \frac{m_k+2}{2m_k^2}\biggr)} .$$
Thus it is easy to identify an infinite number of isospectral collisions for sets of three points.

In a more general setting of an arbitrary number of point masses, we can demonstrate that our optimization is nonconvex.
We show this by using a second partial derivative test.
Consider a squared L2 loss function
$$\mathcal{L}(\mathcal{S} \vert \boldsymbol{I_C}) = \vert \vert \theta_C(\mathcal{S}), \boldsymbol{I_C} \vert \vert_2^2$$
where $\theta_C$ is a function which takes a structure and returns the ordered Cartesian-oriented inertial parameters described above.
For convenience, suppose $\theta_C(\mathcal{S})=(\theta_{x,x},...,\theta_{z,z})$.
We show by the second-derivative test that the optimization problem is nonconvex.
Since the Hessian matrix of second partial derivatives must be positive semi-definite in order for the problem to be convex, it suffices to demonstrate that one term of the Hessian could be negative.
First, for some Cartesian points $(x_i,y_i,z_i),(x_j,y_j,z_j)\in\mathcal{S}$, consider
%$$\frac{\partial^2 \mathcal{L}(\mathcal{S}\vert \boldsymbol{I_C})}{\partial x_i \partial y_i} = 2\biggr[(I_{x,y}-\theta_{x,y})\frac{\partial^2 \theta_{x,y}}{\partial x_i \partial y_j}  -\frac{\partial\theta_{x,y}}{\partial x_i}\frac{\partial\theta_{x,y}}{\partial y_j}$$
%$$ + (I_{z,z}-\theta_{z,z})\frac{\partial^2 \theta_{z,z}}{\partial x_i \partial y_j}-\frac{\partial\theta_{z,z}}{\partial x_i}\frac{\partial\theta_{z,z}}{\partial y_j} \biggr].$$
$$\frac{\partial^2 \mathcal{L}(\mathcal{S}\vert \boldsymbol{I_C})}{\partial x_i \partial y_j} = 2\biggr[(\theta_{x,y}-I_{x,y})\frac{\partial^2 \theta_{x,y}}{\partial x_i \partial y_j}  +\frac{\partial\theta_{x,y}}{\partial x_i}\frac{\partial\theta_{x,y}}{\partial y_j} + (\theta_{z,z}-I_{z,z})\frac{\partial^2 \theta_{z,z}}{\partial x_i \partial y_j}+\frac{\partial\theta_{z,z}}{\partial x_i}\frac{\partial\theta_{z,z}}{\partial y_j} \biggr].$$
We can show the following first-order partials:
%$$\frac{\partial \theta_{x,y}}{\partial x_i} = -m_i(y_i-\bar{y});\; \frac{\partial \theta_{x,y}}{\partial y_j} = -m_j(x_j-\bar{x})$$
%$$\frac{\partial \theta_{z,z}}{\partial x_i} = 2m_i(x_i-\bar{x});\; \frac{\partial \theta_{z,z}}{\partial y_j} = 2m_j(y_j-\bar{y}).$$
$$\frac{\partial \theta_{x,y}}{\partial x_i} = -m_i(y_i-\bar{y});\; \frac{\partial \theta_{x,y}}{\partial y_j} = -m_j(x_j-\bar{x});\;\frac{\partial \theta_{z,z}}{\partial x_i} = 2m_i(x_i-\bar{x});\; \frac{\partial \theta_{z,z}}{\partial y_j} = 2m_j(y_j-\bar{y}).$$
We can also show the following second-order partials:
$$\frac{\partial^2 \theta_{x,y}}{\partial x_i \partial y_j} = \frac{m_i m_j}{\bar{m}};\; \frac{\partial^2 \theta_{z,z}}{\partial x_i \partial y_j} = 0.$$
Then we can substitute these terms into the second-order partial loss function:
$$\frac{\partial^2 \mathcal{L}(\mathcal{S} \vert \boldsymbol{I_C})}{\partial x_i \partial y_i} = 2m_i m_j [ \bar{m}^{-1}(\theta_{x,y}-I_{x,y}) +(x_j-\bar{x})(y_i-\bar{y}) + 4(x_i - \bar{x})(y_j - \bar{y})].$$
Since all three terms may be positive or negative depending on $x_i,x_j,y_i,y_j,\theta_{x,y}$, the Hessian of $\mathcal{L}(\mathcal{S} \vert \boldsymbol{I_C})$ is not positive semi-definite and the optimization is nonconvex.
Since the Jacobian and the Hessian may both be analytically computed (as demonstrated above), we use the BFGS algorithm to minimize the squared L2 loss function \cite{fletcher2000practical}.

\section{Results of Constrained and Unconstrained Environments}\label{append:opt_performance}

\begin{figure}[b]
\includegraphics[width=0.85\textwidth]{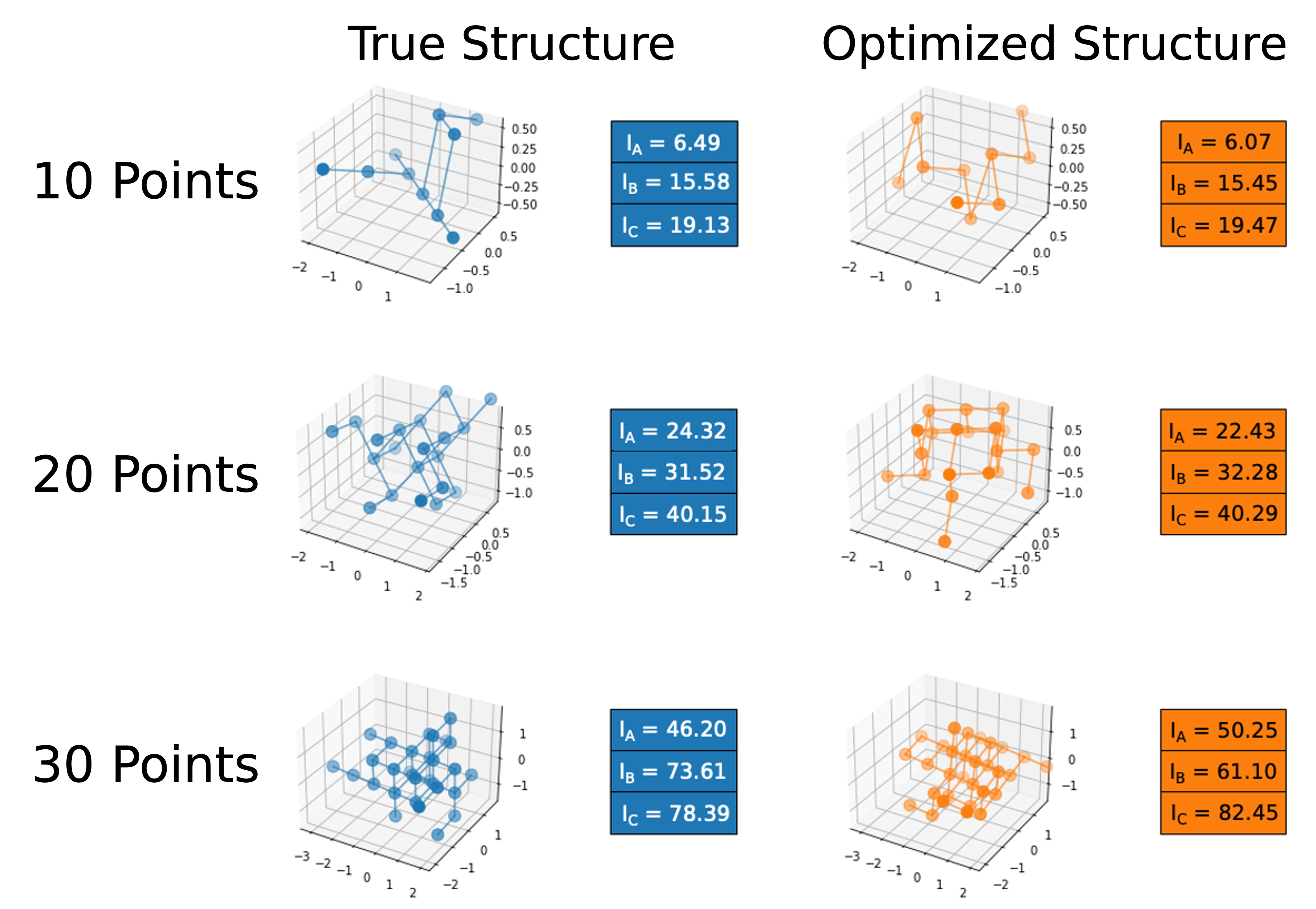}
\caption{\label{fig:isospec_knapsack} Three examples of twins in the constrained environment, with 10, 20, and 30 point masses.
The left-hand side shows the true starting lattice structure, with corresponding moments of inertia $(I_A,I_B,I_C)$.
The right-hand side shows the optimized lattice structure identified using a greedy optimization strategy, with corresponding moments of inertia $(I_A,I_B,I_C)$, optimized to be close to the moments of inertia of the true structure.}
\end{figure}

\autoref{fig:isospec_knapsack} shows three examples of a near-isospectral collision (or twins) in the constrained environment (with 10, 20, and 30 point masses, respectively), as identified by using our iterative greedy approach and oriented according to the same principal rotation axes.
The examples present a trend of how twins identified by a greedy additive approach become harder to generate as we increase the number of point masses.
This can be attributed to the combinatorial explosion of possible structures for the given set of point masses.
%\autoref{fig:constrained_opt_performance} shows the L2 error of nearest collisions for a varying number of point masses, with 1000 tests per number of point masses and 100 restarts per test.
%Our analysis never uncovered an isospectral pair of any size (that was not isomorphic with respect to a translation and/or reflection) in the constrained environment, and it is unclear whether such an isospectral pair could be constructed (either from structures in $\mathbb{R}^3$, or in $\mathbb{R}^n;\; n\geq 2$).

\begin{figure}[b]
\includegraphics[width=0.65\textwidth]{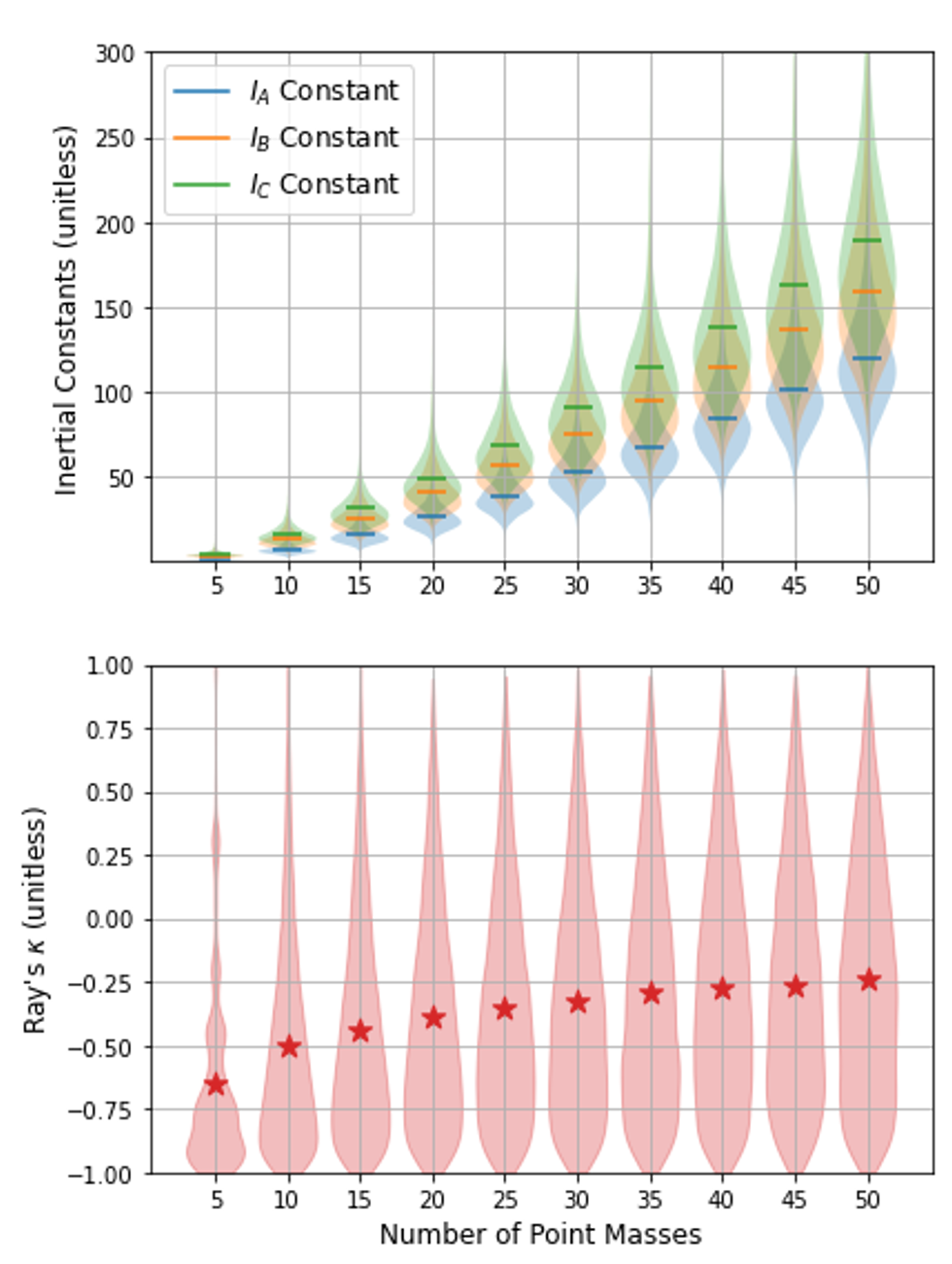}
\caption{\label{fig:point_mass_random} Top: Distribution of moments of inertia for five to 50 point masses.
Bottom: Distribution of Ray's $\kappa$ for five to 50 point masses.}
\end{figure}

We can also consider the distribution of structural geometries among randomly generated structures.
\autoref{fig:point_mass_random} shows the distribution of moments of inertia across 10,000 randomly generated structures of sizes ranging from five to 50 point masses.
%The performance and residual error of the greedy optimization strategy is detailed in Appendix \ref{append:opt_performance}.
As the number of point masses increases, the distribution of moments of inertia widens.
However, the lower plot shows that Ray's asymmetry parameter plateaus at $\kappa \approx -0.2$.
It appears that prolate structures ($\kappa<0$) are far more likely among random geometries in the constrained environment, irrespective how many point masses are added.

\begin{figure}[b]
\includegraphics[width=0.75\textwidth]{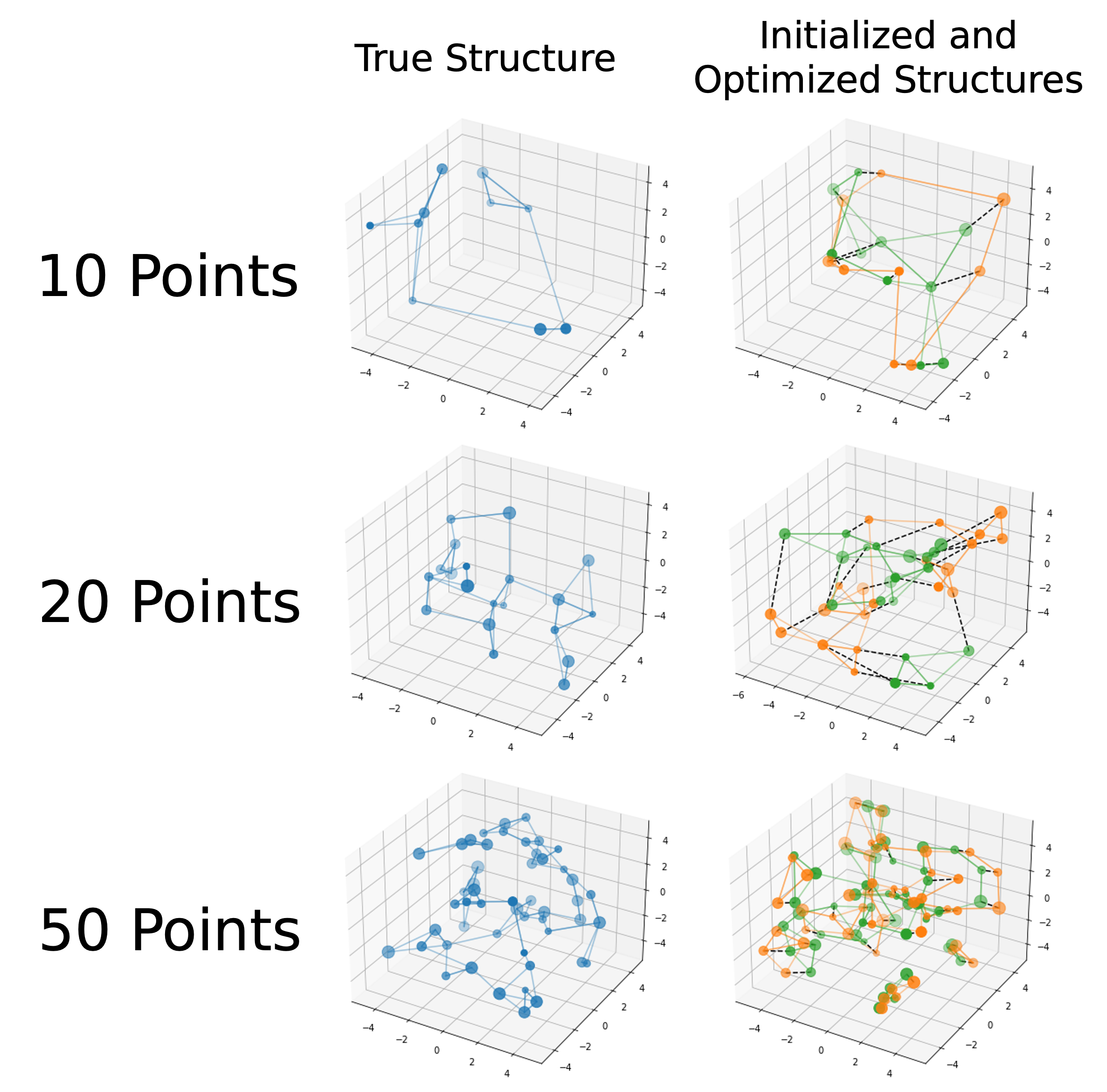}
\caption{\label{fig:isospec_optimization} Isospectral collisions in an unconstrained environment with 10, 20, and 50 point masses.
The left column shows a true random structure to match (blue), while the right column shows the initial random structure (orange) and the final structure (green) which is an isospectral collision with the true random structure (blue).
Black dotted lines indicate the distance covered during the optimization from the initial structure (orange) to the final structure (green).}
\end{figure}

Compared to the constrained environment, generating twins to arbitrary numerical precision is straightforward in the unconstrained environment.
\autoref{fig:isospec_optimization} shows three examples of isospectral collisions with 10, 20, and 50 points identified via our BFGS optimization approach.
We also find that structures are not required to have the same number of point masses to identify isospectral collisions in an unconstrained environment, so long as the number of point masses exceeds three.
The efficiency of the optimization routine for a varying number of point masses is considered in Supplementary Information Section \ref{append:opt_performance}.
Furthermore, an arbitrary number of distinct isospectral collisions can be achieved through this optimization approach. %%%%%%%

\begin{figure}[b]
\includegraphics[width=0.65\textwidth]{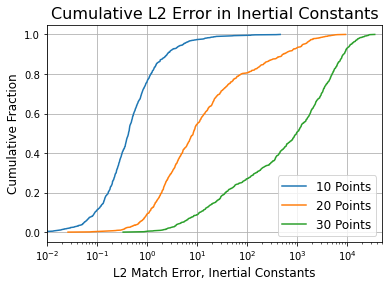}
\caption{\label{fig:constrained_opt_performance} Ordered match error across 1000 independent tests, each with 100 restarts, assessed for 10, 20, and 30 point masses.}
%\ian{Maybe remove heading and include in caption?}
\end{figure}

Next we detail the greedy optimization performance in the constrained environment. \autoref{fig:constrained_opt_performance} shows the L2 error of nearest collisions for a varying number of point masses, with 1000 tests per number of point masses and 100 restarts per test.
We see that the number of points greatly increases the final L2 error associated with the match identified via the greedy optimization procedure.

\begin{figure}[b]
\includegraphics[width=0.65\textwidth]{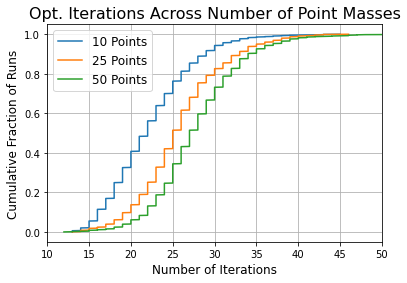}
\caption{\label{fig:unconstrained_opt_performance} Ordered number of iterations required to identify an isospectral collision (within a tolerance of $1\times 10^{-4}$), assessed for 10, 25, and 50 point masses.
}
\end{figure}

Next we detail the BFGS optimization performance in the unconstrained environment.
\autoref{fig:unconstrained_opt_performance} shows the cumulative number of BFGS optimization iterations required to identify an isospectral collision to within $1\times 10^{-4}$ for a varying number of point masses.
We see that $>95\%$ of random structures in the unconstrained environment can be matched to a distinct structure within forty iterations, with fewer optimization iterations required when working with fewer point masses.

\section{PubChem Molecule Analysis}\label{append:pubchem_molecules}
\begin{figure}[b]
\includegraphics[width=0.45\textwidth]{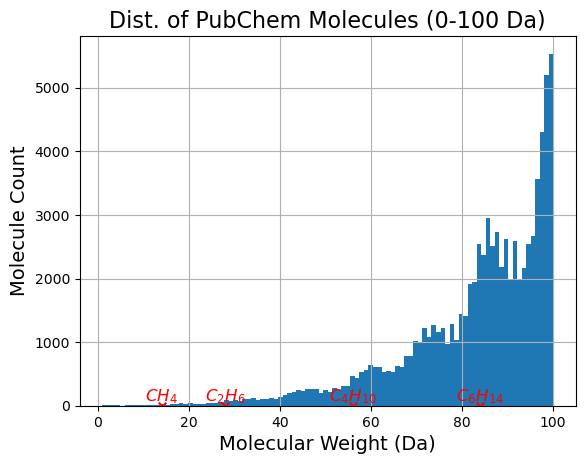}
\includegraphics[width=0.45\textwidth]{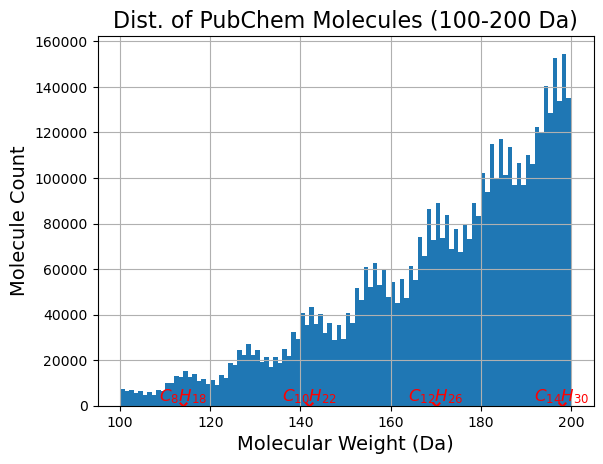}
\includegraphics[width=0.45\textwidth]{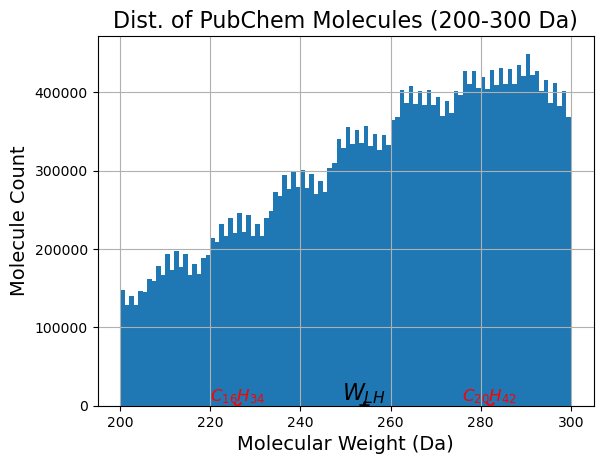}
\includegraphics[width=0.45\textwidth]{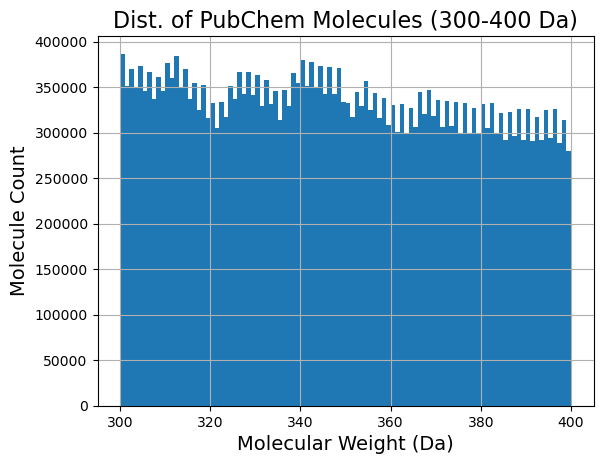}
\includegraphics[width=0.45\textwidth]{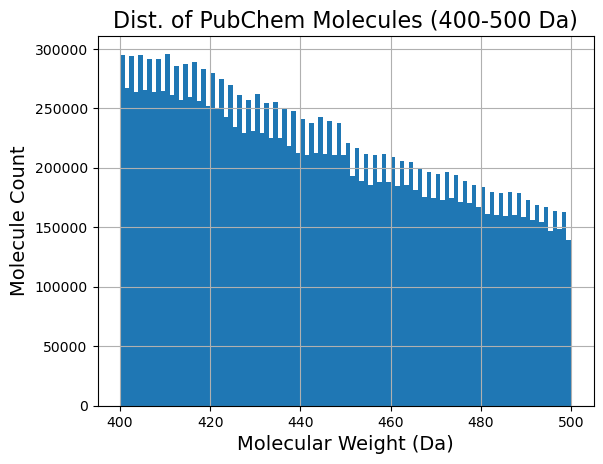}
\includegraphics[width=0.45\textwidth]{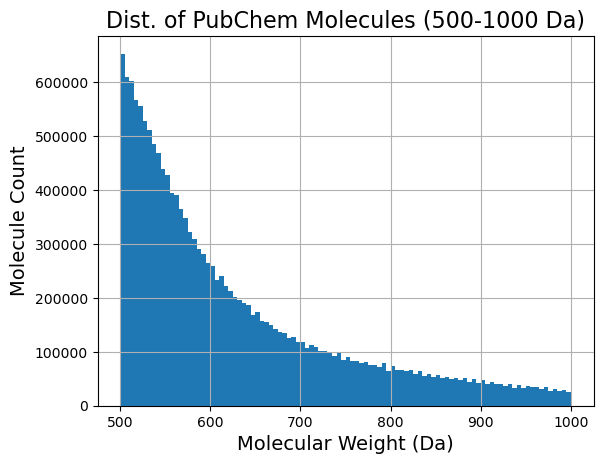}
\includegraphics[width=0.45\textwidth]{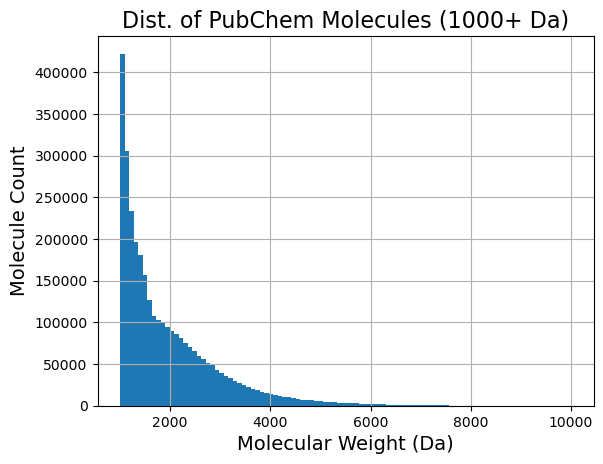}
\caption{\label{fig:pubchem_MW_ALL} Distribution of molecules in PubChem by molecular weight.
These are binned in ranges from 0--100, 100--200, 200--300, 300--400, 400--500, 500--1000, and 1000+ Daltons.}
\end{figure}

In \autoref{fig:pubchem_MW_ALL}, we include histograms across molecular weight (in Da) for all molecules in PubChem, delimited as shown in Figure 4 in the main body of the paper.

\section{QM9 Multi-Fidelity Assessment}\label{append:qm9_multifi}
\begin{figure}[b]
\includegraphics[width=0.65\textwidth]{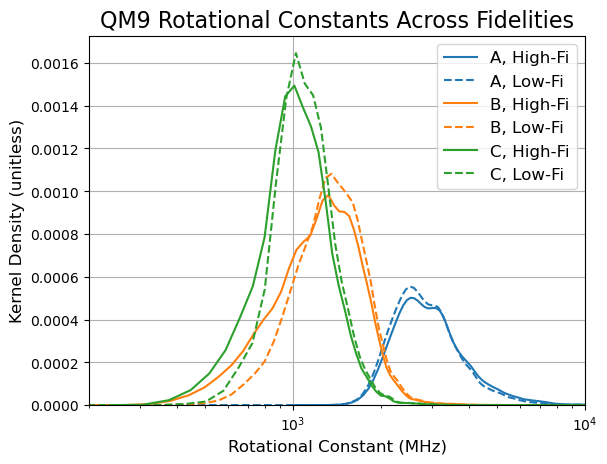}
\caption{\label{fig:qm9_multifi_roto_consts} Distribution of rotational constants on QM9, with geometries assessed by using B3LYP (high-fidelity) and XTB-GFN2 (low-fidelity).}
\end{figure}

\begin{figure}[b]
\includegraphics[width=0.65\textwidth]{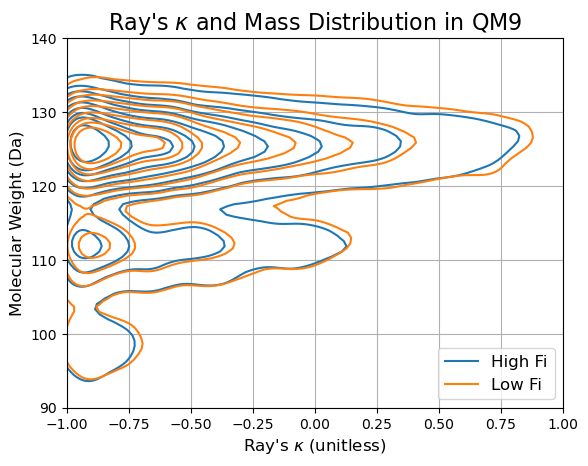}
\caption{\label{fig:qm9_multifi_kappa} Kernel density plot of Ray's $\kappa$ on QM9 across varying molecular weights, with geometries assessed by using B3LYP (high-fidelity) and XTB-GFN2 (low-fidelity).}
\end{figure}

We compared the high-fidelity geometries of QM9 derived via B3LYP/6-31G(2df,p) versus the low-fidelity geometries of QM9 derived via XTB-GFN2.
\autoref{fig:qm9_multifi_roto_consts} shows that the higher-fidelity geometries tend to have greater values for $A$, but lesser values for $B$ and $C$.
\autoref{fig:qm9_multifi_kappa} shows this effect on Ray's $\kappa$, indicating that higher-fidelity measurements are, across all molecular weights, more prolate.

%\nocite{*}

\bibliography{aipsamp}% Produces the bibliography via BibTeX.